\documentclass[journal, twocolumn, 10pt, letterpaper]{IEEEtran}
\usepackage[colorlinks]{hyperref}

\usepackage{siunitx,booktabs}
\usepackage[sort,compress]{cite}
\usepackage{graphicx}
\usepackage{caption}
\usepackage{subcaption}

\usepackage[inline]{enumitem}

\usepackage[linesnumbered,vlined,ruled,norelsize,algo2e]{algorithm2e}

\usepackage[cmex10]{mathtools} 
\usepackage{dsfont,amssymb,bm}
\DeclarePairedDelimiter{\ceil}{\lceil}{\rceil}

\usepackage[amsmath,thmmarks] {ntheorem}
\newtheorem {problem}       {Problem}
\newtheorem {property}      {Property}

\newtheorem {theorem}{Theorem}
\newtheorem{corollary}{Corollary}
\newtheorem {proposition}{Proposition}
\newtheorem {definition}{Definition}
\newtheorem {lemma}{Lemma}

\newtheorem {assumption}{Assumption}

\theoremstyle{plain}
\theorembodyfont{\normalfont}
\newtheorem {remark}{Remark}

\theoremstyle{plain}
\theoremheaderfont{\normalfont\itshape}
\theorembodyfont{\normalfont}

\theoremstyle{plain}
\theoremheaderfont{\normalfont\itshape}
\theorembodyfont{\normalfont}
\newtheorem *{proof}{Proof:}

\usepackage[T1]{fontenc}

\newcommand\VEC{\bm}
\newcommand\ALPHABET{\mathbb}
\newcommand\BLANK{\mathfrak {E}}
\newcommand\ASU{\mathrm{SU}}
\newcommand\SQC{\mathrm{SQC}}

\newcommand\IND{\mathds{1}}
\newcommand\PR {\mathds{P}}
\newcommand\EXP{\mathds{E}}

\newcommand\reals{\mathds{R}}
\newcommand\integers{\mathds{Z}}

\newcommand\DEFINED{\coloneqq}

\newcommand{\JL}{\mathsf{L}}
\newcommand{\M}{\mathsf{M}}

\newcommand{\N}{\mathsf{N}}
\newcommand{\C}{\mathsf{C}}

\begin{document}

\title {Remote estimation over a packet-drop channel with Markovian state}
\author{Jhelum Chakravorty and Aditya Mahajan%
\thanks{Preliminary version of this paper was presented in the 2016 IFAC Workshop on Distributed Estimation and Control in Networked Systems (NecSys), in 2017 International Symposium of Information Theory (ISIT) and in 2017 American Control Conference (ACC).}%
\thanks{The authors are with the Department of Electrical and Computer Engineering, McGill University, QC, Canada. Email: \texttt{\{jhelum.chakravorty@mail., aditya.mahajan@\}mcgill.ca}.
This research was funded through NSERC Discovery Accelerator Grant 493011.}%
}

\maketitle

\begin{abstract}
  We investigate a remote estimation problem in which a transmitter observes a
  Markov source and chooses the power level to transmit it over a time-varying
  packet-drop channel. The channel is modeled as a channel with Markovian
  state where the packet drop probability depends on the channel state and the
  transmit power. A receiver observes the channel output and the channel state
  and estimates the source realization. The receiver also feeds back the
  channel state and an acknowledgment for successful reception to the
  transmitter. We consider two models for the source---finite state Markov
  chains and first-order autoregressive processes. For the first model, using
  ideas from team theory, we establish the structure of optimal transmission
  and estimation strategies and identify a dynamic program to determine
  optimal strategies with that structure. For the second model, we assume that
  the noise process has unimodal and symmetric distribution. Using ideas from
  majorization theory, we show that the optimal transmission strategy is
  symmetric and monotonic and the optimal estimation strategy is like Kalman
  filter. Consequently, when there are a finite number of power levels, the
  optimal transmission strategy may be described using thresholds that depend
  on the channel state. Finally, we propose a simulation based approach
  (Renewal Monte Carlo) to compute the optimal thresholds and optimal
  performance and elucidate the algorithm with an example. 
\end{abstract}


\section{Introduction}

\subsection{Motivation and literature overview}

Network control systems are distributed systems where plants, sensors,
controllers, and actuators are interconnected via a communication network.
Such systems arise in a variety of applications such as IoT (Internet of
Things), smart grids, vehicular networks, robotics, etc. One of the
fundamental problem in network control system is \emph{remote
estimation}---how should a sensor (which observes a stochastic process)
transmit its observations to a receiver (which estimates the state of the
stochastic process) when there is a constraint on communication, either in
terms of communication cost or communication rate. 

\begin{figure}[!t]
  \centering
  \includegraphics[width=\linewidth]{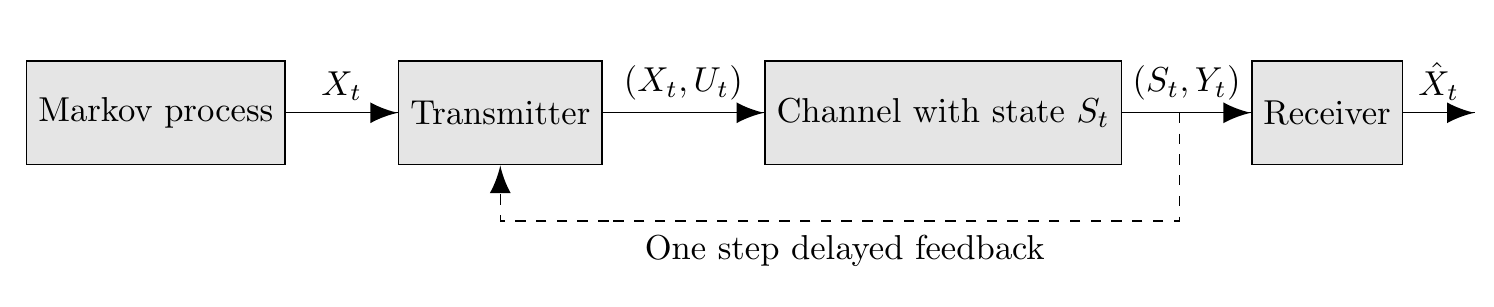}
  \caption{Remote estimation over channel with state.}
  \label{fig:block}
\end{figure}

In this paper, we consider a remote estimation system as shown in
Fig.~\ref{fig:block}. The system consists of a sensor and an estimator
connected over a time-varying wireless fading channel. The sensor observes a
Markov process and chooses the power level to transmit its observation to the
remote estimator. Communication is noisy and the transmitted packet may get
dropped according to a probability that depends on the channel state and the
power level. When the packet is dropped the receiver generates an estimate of
the state of the source according to previously received packets. The
objective is to choose power control and estimation strategies to minimize a
weighted sum of transmission power and estimation error. 

Several variations of the above model have been considered in the literature.
Models with noiseless communication channels have been considered
in~\cite{XuHes2004a,ImerBasar,Rabi2012,LipsaMartins:2011,NayyarBasarTeneketzisVeeravalli:2013,MH2017}.
Since the channel is noiseless, these papers assume that there are only two
power levels: power level 0, which corresponds to not transmitting; and power
level 1, which corresponds to transmitting. Under slightly different modeling
assumptions, these papers identify the structure of optimal transmission and
estimation strategies for first-order autoregressive sources with unimodal
noise and for higher order autoregressive sources with orthogonal dynamics and
isotropic Gaussian noise. It is shown that the optimal transmission strategy
is threshold-based, i.e., the sensor transmits whenever the current
\emph{error} is greater than a threshold. It is also shown that the optimal
estimation strategy is like Kalman filter: when the receiver receives a packet, the
estimate is the received symbol; when it does not receive the packet, then the
estimate is the one-step prediction based on the previous symbol. Quite
surprisingly, these results show that there is no advantage in trying to
extract information about the source realization from the choice of the power
levels. The transmission strategy at the sensor is also called
\emph{event-triggered} communication because the sensor transmits when the
event `error is greater than a threshold' is triggered. 
%In addition, different versions of remote estimation with noisy channels have been considered in the literature.
 Models with i.i.d.\@ packet-drop channels are considered in~\cite{JC_AM_IFAC16,LipsaMArtins2009,JC-JS-AM-ACC17}, where it is assumed that the transmitter has two power levels: \textsc{on} or \textsc{off}. Remote estimation over additive noise channel is considered in~\cite{gaoEtal18}. 

In this paper we consider a remote estimation problem over packet-drop channel with
Markovian state. We assume that the receiver observes the channel state and
feeds it back to the transmitter with one step delay. Preliminary results for
this model are presented in~\cite{JC_AM_ISIT17}, where attention was
restricted to a binary state channel with two input power values (ON or OFF).
In the current paper, we consider arbitrary number of channel states and power
levels. A related paper is~\cite{Renetal2017}, in which a remote estimation
over packet-drop channels with Markovian state is considered. It is
assumed that the sensor and the receiver know the channel state. It is shown
that optimal estimation strategies are like Kalman filter. A detailed comparison
with~\cite{Renetal2017} is presented in Section~\ref{sec:compare-ren}. 

Several approaches for computing the optimal transmission strategies have been
proposed in the literature. For noiseless channels, these include dynamic
programing based approaches~\cite{LipsaMartins:2011,
NayyarBasarTeneketzisVeeravalli:2013, MH2012}, approximate dynamic programming
based approaches~\cite{Gatsisetal2014}, renewal theory based
approaches~\cite{JC_AM_TAC17}. It is shown in~\cite{Heetal} that for
event-triggered scheduling, the posterior density follows a generalized closed
skew normal (GCSN) distribution. For Markovian channels (when the state is not
observed), a change of measure technique to evaluate the performance of an
event-triggered scheme is presented in~\cite{Chenetal2017}. In this paper, we
present a renewal theory based Monte Carlo approach for computing the optimal thresholds. A preliminary version of the results was presented
in~\cite{JC-JS-AM-ACC17} for a channel with i.i.d.\@ packet drops. 

\subsection{Contributions}
In this paper, we investigate team optimal transmission and estimation
strategies for remote estimation over time varying packet-drop channels. We
consider two models for the source: finite state Markov source and first order
autoregressive source (over either integers or reals). Our main contributions
are as follows.
\begin{enumerate}
  \item For finite sources, we identify sufficient statistics for both the
    transmitter and the receiver and obtain a dynamic programming
    decomposition to compute optimal transmission and estimation strategies.
  \item For autoregressive sources, we identify qualitative properties of
    optimal transmission and estimation strategies. In particular, we show
    that the optimal estimation strategy is like Kalman filter and the optimal
    transmission strategy only depends on the current source realization and
    the previous channel state (and does not depend on the receiver's belief
    of the source). Furthermore, when the channel state is stochastically
    monotone (see Assumption~\ref{assump:sto-mono} for definition), then for
    any value of the channel state, the optimal transmission strategy is
    symmetric and quasi-convex in the source realization. Consequently, when
    the power levels are finite, the optimal transmission strategy is
    threshold-based, where the thresholds only depend on the previous channel
    state.
  \item We show that the above qualitative properties extend naturally to
    infinite horizon models. 
  \item For infinite horizon models, we present a Renewal Theory based
    Monte-Carlo algorithm to evaluate the performance of any threshold-based
    strategy. We then combine it with a simultaneous perturbation based
    stochastic approximation algorithm to provide an algorithm to compute the
    optimal thresholds. We illustrate our results with a numerical example of
    a remote estimation problem with a transmitter with two power levels and a
    Gilbert-Elliott erasure channel.  
  \item We show that the problem of transmitting over one of $m$ available
    i.i.d.\@ packet-drop channels (at a constant power level) can be considered
    as special case of our model. We show that there exist thresholds
    $\{k^{(i)}_t\}_{i=1}^m$, such that it is optimal to transmit over channel
    $i$ if the \emph{error state} $E_t \in [k^{(i)}_t, k^{(i+1)}_t)$. See
    Sec.~\ref{subsec:multiple_iid_channel} for details. 
\end{enumerate}

\subsection{Notation}
We use uppercase letters to denote random variables (e.g, $X$, $Y$, etc),
lowercase letters to denote their realizations (e.g., $x$, $y$, etc.).
$\integers$, $\integers_{\ge 0}$ and $\integers_{> 0}$ denote respectively the
sets of integers, of non-negative integers and of positive integers.
Similarly, $\reals$, $\reals_{\ge 0}$ and $\reals_{> 0}$ denote respectively
the sets of reals, of non-negative reals and of positive reals. For any set
$\ALPHABET A$, let $\IND_{\ALPHABET A}$ denote its indicator function, i.e.,
$\IND_{\ALPHABET A}(x)$ is $1$ if $x \in \ALPHABET A$, else $0$. $|\ALPHABET
A|$ denotes the cardinality of set $\ALPHABET A$. $\Delta(\ALPHABET X)$
denotes the space of probability distributions of $\ALPHABET X$. For any
vector $v \in \reals^n$, $v_i \in \reals$ denotes the $i$-th component of $v$.
For any vector $v$ and an interval $A = [a,b]$ of $\reals$, $w = [v]_A$ means
that $w_i$ equals $a$ if $v_i \le a$; equals $v_i$ if $v_i \in (a,b)$; and
equals $b$ if $v_i \ge b$. Given a Borel subset $A \subseteq \reals$ and a
density $\pi$, we use the notation $\pi(A) \DEFINED \int_A \pi(e)de$. For any
vector $\mathbf v$, $\nabla_{\mathbf v}$ denotes the derivative with respect
to $\mathbf v$.

\subsection{The communication system}\label{sec:system-model}
We consider a remote estimation system shown in Fig.~\ref{fig:block}. The
different components of the system are explained below. 

\subsubsection{Source model} 
The source is a first-order time-homogeneous Markov chain $\{X_t\}_{t \ge 0}$,
$X_t \in \ALPHABET X$. We consider two models for the source.
\begin{itemize}
  \item \textbf{Finite state Markov source.} In this model, we assume that
    $\ALPHABET X$ is a finite set and denote the state transition matrix by
    $P$, i.e., for any $x,y \in \ALPHABET X$, $P_{xy} = \PR({X_{t+1} = y}
    \,|\allowbreak\, {X_t = x})$. 
  \item \textbf{First-order autoregressive source.} In this model, we
    assume that $\ALPHABET X$ is either $\ALPHABET Z$ or $\ALPHABET R$. The
    initial state $X_0 = 0$ and for $t \ge 0$, the source evolves as
    \begin{equation}\label{eq:AR1}
      X_{t+1} = a X_t + W_t,
    \end{equation}
    where $a, W_t \in \ALPHABET X$ and $\{W_t\}_{t \ge 0}$ is an i.i.d.\@
    sequence where $W_t$ is distributed according to a symmetric and unimodal
    distribution\footnote{With a slight abuse of notation, when $\ALPHABET X =
      \reals$, we consider $\mu$ to the probability density function and when
      $\ALPHABET X = \integers$, we consider $\mu$ to be the probability mass
    function.} $\mu$. 
\end{itemize}

\subsubsection{Channel model}
The channel is a packet-drop channel with state. The state process $\{S_t\}_{t
\ge 0} \in \ALPHABET S$, is a first-order time-homogeneous Markov chain with
transition probability matrix~$Q$. We assume that $\ALPHABET S$ is finite.
This is a standard model for time-varying wireless
channels~\cite{GoldsmithVaraiya1996,YangEtAl2005}.

The input alphabet of the channel is $\ALPHABET X$ and the output alphabet is
$\ALPHABET Y \DEFINED \ALPHABET X \cup \{\BLANK \}$ where the symbols $\BLANK$
denotes that no packet was received. At time $t$, the channel output is
denoted by~$Y_t$. 

The packet drop probability depends on the input power $U_t \in \ALPHABET U$,
where $\ALPHABET U$ is the set of allowed power levels. We assume that
$\ALPHABET U$ is a subset of $\reals_{\ge 0}$ and $\ALPHABET U$ is either a
finite set of the form $\{0, u_{(1)}, \dots, u_{\max}\}$ or an interval of the
form $[0,u_{\max}]$, i.e., $\ALPHABET U$ is uncountable. When $U_t = 0$, it
means that the transmitter does not send a packet. In particular, for any
realization $( x_{0:T}, s_{0:T}, u_{0:T}, y_{0:T})$ of $(X_{0:T}, S_{0:T},
U_{0:T}, Y_{0:T})$, we have 
\begin{multline}
  \label{eq:state}
  \PR(S_t = s_t \mid X_{0:t} = x_{0:t}, S_{0:t-1} = s_{0:t-1},
  U_{0:t} = u_{0:t}) 
  \\ =
  \PR(S_t = s_t \mid S_{t-1} = s_{t-1}) = Q_{s_{t-1} s_t},
\end{multline}
and
\begin{multline}
  \label{eq:channel}
  \PR(Y_t = y_t \mid X_{0:t} = x_{0:t}, S_{0:t} = s_{0:t}, U_{0:t}
  = u_{0:t}) 
  \\ =
  \begin{cases}
    1-p(s_t,u_t), & \mbox{if $y_t = x_t$} \\
    p(s_t,u_t),   & \text{if $y_t = \BLANK$}\\
    0, & \mbox{otherwise},
  \end{cases}
\end{multline}
where $p(s_t,u_t)$ is the probability that a packet transmitted with power
level $u_t$ when the channel is in state $s_t$ is dropped. We assume that the
set $\ALPHABET S$ of the channel states is an ordered set where a larger state
means a better channel quality. Then, for all $s \in \ALPHABET S$, $p(s,u)$ is
(weakly) decreasing in~$u$ with $p(s,0) = 1$ and $p(s,u_{\max}) \ge 0$.
Furthermore, we assume that for all $u \in \ALPHABET U$, $p(s,u)$ is
decreasing in $s$. 

\subsection{The decision makers and the information structure}

There are two decision makers in the system---the transmitter and the
receiver. At time~$t$, the transmitter chooses the transmit power $U_t$ while
the receiver chooses an estimate $\hat X_t \in \ALPHABET X$. Let $I^1_t$ and
$I^2_t$ denote the information sets at the transmitter and the receiver
respectively. 

The transmitter observes the source realization $X_t$. In addition, there is
one-step delayed feedback from the receiver to the transmitter.\footnote{Note
that feedback of $Y_t$ requires 1 bit to indicate whether the packet was
received or not and feedback of $S_t$ requires $\ceil[\big]{\log_2 |\ALPHABET
S|}$ bits.} Thus, the information available at the transmitter~is
\[
  I^1_t = \{X_{0:t}, U_{0:t-1}, S_{0:t-1}, Y_{0:t-1} \}.
\]
The transmitter chooses the transmit power $U_t$ according to 
\begin{equation}\label{eq:tx-1}
  U_t = f_t(I^1_t) = f_t(X_{0:t}, U_{0:t-1}, S_{0:t-1}, Y_{0:t-1}),
\end{equation}
where $f_t$ is called the \emph{transmission rule} at time~$t$. The collection $\VEC f \DEFINED (f_0, f_1, \dots)$ for all time is called the
\emph{transmission strategy}.

The receiver observes $Y_t$ and, in addition, observes the channel state
$S_t$. Thus, the information available at the receiver~is
\[
  I^2_t = \{S_{0:t}, Y_{0:t} \}.
\]
The receiver chooses the estimate $\hat X_t$ is chosen according to
\begin{equation} \label{eq:rx-1}
  \hat X_t = g_t(I^2_t) = g_t(S_{0:t}, Y_{0:t}),
\end{equation}
where $g_t$ is called the \emph{estimation rule} at time~$t$. The collection
$\VEC g \DEFINED (g_0, g_1, \dots)$ for all time is called the
\emph{estimation strategy}.

The collection $(\VEC f, \VEC g)$ is called a \emph{communication strategy}.

\subsection{The performance measures and problem formulation}

At each time~$t$, the system incurs two costs: a transmission cost
$\lambda(U_t)$ and a distortion or estimation error $d(X_t, \hat X_t)$. Thus, the per-step cost is
\[
  c(X_t, U_t, \hat X_t) = \lambda(U_t) + d(X_t, \hat X_t).
\]
We assume that $\lambda(u)$ is (weakly) increasing in $u$ with $\lambda(0) = 0$ and $\lambda(u_{\max}) < \infty$.
For the autoregressive source model, we assume that the distortion is given by
$d({X_t - \hat X_t})$, where $d(\cdot)$ is even and quasi-convex with $d(0) = 0$. 

We are interested in the following optimization problems:

\begin{problem}[Finite horizon] \label{prob:finite}
  In the model described above, identify a communication strategy $(\VEC f^*,
  \VEC g^*)$ that minimizes the total cost given by
  \begin{equation}
  \label{eq:fin-cost}
    J_T(\VEC f, \VEC g) \DEFINED \EXP\bigg[\sum_{t=0}^{T-1} c(X_t, U_t, \hat X_t) \bigg].
  \end{equation}
\end{problem}

\begin{problem}[Infinite horizon] \label{prob:infinite}
  In the model described above,  given a discount factor $\beta \in (0,1]$, identify a communication strategy $(\VEC f^*, \VEC g^*)$
  that minimizes the total cost given as follows:
  \begin{enumerate}
    \item For $\beta \in (0,1)$, 
      \begin{equation} \label{eq:infin-cost-dis}
        J_\beta(\VEC f, \VEC g) = (1-\beta)
        \EXP\bigg[ \sum_{t=0}^\infty \beta^t c(X_t, U_t, \hat X_t) \bigg].
      \end{equation}
    \item For $\beta = 1$, 
      \begin{equation} \label{eq:infin-cost-avg}
        J_1(\VEC f, \VEC g) = \lim_{T \to \infty} \frac 1T 
        \EXP\bigg[ \sum_{t=0}^{T-1} c(X_t, U_t, \hat X_t) \bigg].
      \end{equation}
  \end{enumerate}
\end{problem}

\begin{remark}\label{rem:coding}
  In the above model, it has been assumed that whenever the transmitter
  transmits (i.e., $U_t \neq 0$), it sends the source realization uncoded.
  This is without loss of generality because the channel input alphabet is the
  same as the source alphabet and the channel is symmetric. For such models,
  coding does not improve performance~\cite{WalrandVaraiya:1983}. 
\end{remark}

Problems~\ref{prob:finite} and~\ref{prob:infinite} are decentralized
stochastic control problems. The main conceptual difficulty in solving such
problems is that the information available to the decision makers and hence
the domain of their strategies grow with time, making the optimization problem
combinatorial. One could circumvent this issue by identifying a suitable
\emph{information state} at the decision makers, which do not grow with time.
In the following section, we discuss one such method to establish the
structural results. 

\section{Main results for finite state Markov sources}

\subsection{Structure of optimal communication strategies}
\label{sec:main-results-finite-MC}

We establish two types of structural results. First, we use
\emph{person-by-person approach} to show that $(X_{0:t-1}, U_{0:t-1})$ is
irrelevant at the transmitter (Lemma~\ref{lemma:irrelevant_info_Tx}); then, we
use the common information approach of~\cite{NMT:partial-history-sharing} and
establish a belief-state for the common information $(S_{0:t}, Y_{0:t})$
between the transmitter and the receiver (Theorem~\ref{thm:structure-fin}). 

\begin{lemma}\label{lemma:irrelevant_info_Tx}
  For any estimation strategy of the form~\eqref{eq:rx-1}, there is no loss
  of optimality in restricting attention to transmission strategies of the
  form
  \begin{equation} \label{eq:tx-2}
    U_t = f_t(X_t, S_{0:t-1}, Y_{0:t-1}).
  \end{equation}
\end{lemma}

The proof proceeds by establishing that the process $\{X_t, S_{0:t-1},
Y_{0:t-1}\}_{t \ge 0}$ is a controlled Markov process controlled by $\{U_t\}_{t \ge 0}$. See Appendix~\ref{app:proof-lemma-irr-info} for details.

For any strategy $\VEC f$ of the form~\eqref{eq:tx-2} and any realization
$(s_{0:T}, y_{0:T})$ of $(S_{0:T}, Y_{0:T})$, define $\varphi_t \colon
\ALPHABET X \to \ALPHABET U$ as
\[
  \varphi_t(x) = f_t(x, s_{0:t-1}, y_{0:t-1}), 
  \quad \forall x \in \ALPHABET X.
\]
Furthermore, define conditional probability measures~$\pi^1_t$ and $\pi^2_t$
on $\ALPHABET X$ as follows: for any $x \in \ALPHABET X$,
\begin{align*}
  \pi^1_t(x) &\DEFINED \PR^{\VEC f}(X_t = x \mid 
    S_{0:t-1} = s_{0:t-1}, Y_{0:t-1} = y_{0:t-1} ), 
  \\
  \pi^2_t(x) &\DEFINED \PR^{\VEC f}(X_t = x \mid 
  S_{0:t} = s_{0:t}, Y_{0:t} = y_{0:t} ).
\end{align*}
We call $\pi^1_t$ the \emph{pre-transmission belief} and $\pi^2$ the
\emph{post-transmission belief}. 
Note that when $(S_{0:T}, Y_{0:T})$ are random variables, then $\pi^1_t$ and
$\pi^2_t$ are also random variables (taking values in $\Delta(\ALPHABET X)$),
which we denote by $\Pi^1_t$ and $\Pi^2_t$.

For the ease of notation, define $B(\pi^1, s, \varphi)$ as follows:
\begin{align}
  B(\pi^1, s_t, \varphi) &\DEFINED 
  \PR (Y_t = \BLANK\,|\, S_{0:t} = s_{0:t}, Y_{0:t-1}=y_{0:t-1}) \notag\\
  &= \sum_{x_t \in \ALPHABET X} \pi^1(x_t) p(s_t,\varphi(x_t)). \label{eq:B01}
\end{align}
Furthermore, define $\pi^1|_{\varphi_t,s}$ as follows:
\begin{equation}
  \pi^1|_{\varphi,s}(x) \DEFINED \frac{\pi^1(x) p(s,\varphi(x))}{B(\pi^1, s, \varphi)}.
  \label{eq:pi}
\end{equation}

Then, using Baye's rule one can show the following:
\begin{lemma}\label{lemma:F1-F2}
  Given any transmission strategy $\VEC f$ of the
  form~\eqref{eq:tx-2}:
  \begin{enumerate}
    \item there exists a function~$F^1$ such that
      \begin{equation}
        \pi^1_{t+1} = F^1(\pi^2_t) = \pi^2_t P.
        \label{eq:F1} 
      \end{equation}
    \item there exists a function $F^2$ such that
      \begin{equation}
        \pi^2_t = F^2(\pi^1_t, s_{t}, \varphi_{t}, y_t)
        = \begin{cases}
          \delta_{y_t}, & \mbox{if $y_t \in \ALPHABET X$} 
          \\
          \pi^1_t|_{\varphi_t,s_{t}}, & \mbox{if $y_t = \BLANK$}.
        \end{cases}
        \label{eq:F2}
      \end{equation}
  \end{enumerate}
\end{lemma}

Note that in~\eqref{eq:F1}, we are treating $\pi^2_t$ as a
row-vector and in~\eqref{eq:F2}, $\delta_{y_t}$ denotes a Dirac measure
centered at $y_t$. The update equations~\eqref{eq:F1} and
\eqref{eq:F2} are standard non-linear filtering equations.
See supplementary material for proof.

\begin{theorem} \label{thm:structure-fin}
  In Problem~\ref{prob:finite} with finite state Markov source, we have that:
  \begin{enumerate} 
    \item \emph{Structure of optimal strategies:}
      There is no loss of optimality in restricting attention to transmission
      and estimation strategies of the form:
      \begin{align}
        U_t &= f^*_t(X_t, S_{t-1}, \Pi^1_t), 
        \label{eq:tx-*-fin}\\
        \hat X_t &= g^*_t(\Pi^2_t).
        \label{eq:rx-*-fin}
      \end{align}
    \item \emph{Dynamic program:} Let $\Delta(\ALPHABET X)$ denote the space
      of probability distributions on $\ALPHABET X$. Define value functions
      $V^1_t \colon \Delta(\ALPHABET X) \times \ALPHABET S \to \reals$ and
      $V^2_t \colon \Delta(\ALPHABET X) \times \ALPHABET S \to \reals$ as
      follows: for any $s_t \in \ALPHABET S$,
      \begin{equation}
        V^1_{T+1}(\pi^1_t, s_t) = 0, 
      \end{equation}
      and for $t \in \{T, \dots, 0\}$
      \begin{align}
        V^1_t(\pi^1_t,s_t) &= 
        \min_{\varphi_t \colon \ALPHABET X \to \ALPHABET U} 
        \big\{ 
          \Lambda (\pi^1_t, \varphi_t) + H_t(x_t, \pi^1_t, s_t, \varphi_t) 
        \big\},
        \label{eq:dp-1-fin}
        \\
        V^2_t(\pi^2_t,s_t) &=
        \min_{\hat x \in \ALPHABET X} D(\pi^2_t, \hat x) 
        + V^1_{t+1}(\pi^2_t P, s_t),
        \label{eq:dp-2-fin}
      \end{align}
      where 
      \begin{align*}
        \Lambda (\pi^1, \varphi) &\DEFINED 
        \sum_{x \in \ALPHABET X} \lambda(\varphi(x)) \pi^1(x), \\
        H_t(x,\pi^1, s, \varphi) & \DEFINED
        B(\pi^1, s, \varphi) V^2_t(\delta_{x},s)
        \\
        & \qquad + (1-B(\pi^1, s, \varphi)) V^2_t(\pi^1|_{\varphi,s},s),
        \\
        D(\pi^2, \hat x) & \DEFINED 
        \sum_{x \in \ALPHABET X} d(x, \hat x) \pi^2(x).
      \end{align*}

      Let $\Psi_t(s, \pi^1)$ denote the arg min of the right hand side
      of~\eqref{eq:dp-1-fin} and $g^*_t(\pi^2_t) \DEFINED \arg\min_{\hat x \in
      \ALPHABET X} D(\pi^2, \hat x)$. Then, the optimal transmission strategy
      is given by
      \[
        f^*_t(\cdot, s, \pi^1_t) = \Psi_t(s, \pi^1_t)
      \]
      and the optimal estimation strategy is given by $g^*_t$.
  \end{enumerate}
\end{theorem}
The proof follows from the \emph{common information
approach}~\cite{NMT:partial-history-sharing}. See
Appendix~\ref{app:proof_thm-stru_fin} for details.

\begin{remark}
  The first term in~\eqref{eq:dp-1-fin} is the expected communication cost,
  the second term is the expected cost-to-go. The first term
  in~\eqref{eq:dp-2-fin} is the expected distortion and the second term is the
  expected cost-to-go.
\end{remark}

\begin{remark}\label{rem:compact-U}
  In~\eqref{eq:dp-1-fin} we use $\min$ instead of $\inf$ for the following
  reasons. Let $\Phi$ denote the set of functions from $\ALPHABET X$ to
  $\ALPHABET U$, which is equal to $\prod_{x \in \ALPHABET X} \ALPHABET U$
  (since $\ALPHABET X$ is finite). When $\ALPHABET U$ is finite, $\Phi$ is
  also finite and thus we can use $\min$ in~\eqref{eq:dp-1-fin}. When
  $\ALPHABET U$ is uncountable, $\Phi$ is  a product of compact sets and hence
  is compact and thus we can use $\min$ in~\eqref{eq:dp-1-fin}. 
\end{remark}

\begin{remark}\label{rem:infinite-hor}
  Note that the dynamic program in Theorem~\ref{thm:structure-fin} is similar
  to a dynamic program for a partially observable Markov Decision Process
  (POMDP) with finite state space and finite or uncountable action space (see
  Remark~\ref{rem:compact-U}). Thus, the dynamic program can be extended to
  infinite horizon discounted cost model after verifying standard assumptions.
  However, doing so does not provide any additional insight, so we do not
  present infinite horizon results for this model. We will do so for the
  autoregressive source model later in the paper, where we provide an
  algorithm to find the optimal time-homogeneous strategy for infinite horizon
  criteria.  
\end{remark}

\section{Main results for autoregressive sources}\label{sec:fin-AR}

\subsection{Structure of optimal trategies for finite horizon model}
\label{sec:error_process}

We start with a change of variables. Define a process $\{Z_t\}_{t \ge 0}$ as follows: $Z_0 = 0$ and for $t \ge 0$,
\[
  Z_{t} = \begin{cases}
    a Z_{t-1}, & \mbox{if $Y_t = \BLANK$} \\
    Y_t, & \mbox{if $Y_t \in \ALPHABET X$}. 
  \end{cases}
\]
 
Next, define processes $\{E_t\}_{t \ge 0}$, $\{E^+_t\}_{t \ge 0}$, which we
call the \emph{error processes} and $\{\hat E_t\}_{t \ge 0}$ as follows:
\[
  E_t \DEFINED X_t - a Z_{t-1}, \quad
  E^+_t \DEFINED X_t - Z_t, \quad
  \hat E_t \DEFINED \hat X_t - Z_t.
\]
The processes $\{E_t\}_{t \ge 0}$ and $\{ E^+_t \}_{t \ge 0}$ are related
as follows: $E_0 = 0$, $E^+_0 = 0$, and for $t \ge 0$,
\begin{equation} \label{eq:e-update}
  E^+_t = \begin{cases}
    E_t, & \mbox{if $Y_t = \BLANK$} \\
    0  , & \mbox{if $Y_t \in \ALPHABET X$}
  \end{cases}
  \quad \text{and} \quad
  E_{t+1} = a E^+_t + W_t.
\end{equation}
The above dynamics may be rewritten as
\begin{equation}\label{eq:dynamics-2a}
  E_{t+1} = \begin{cases}
    a E_t + W_t, & \hbox{if $Y_t = \BLANK$} \\
    W_t, & \hbox{if $Y_t \neq \BLANK$}.
  \end{cases}
\end{equation}

Since $X_t - \hat X_t = E^+_t - \hat E_t$, we have that $d(X_t - \hat X_t) =
d(E^+_t - \hat E_t)$. Thus, with this change of variables, the per-step cost
may be written as $\lambda(U_t) + d(E^+_t - \hat E_t)$.

Note that $Z_t$ is a deterministic function of $Y_{0:t}$. Hence, at time $t$, $Z_{t-1}$ is measurable at the transmitter and thus $E_t$ is measurable at the transmitter. Moreover, at time $t$, $Z_t$ is measurable at the receiver. 

\begin{lemma}\label{lemma:U_hat_E}
  For any transmission and estimation strategies of the form~\eqref{eq:tx-2}
  and~\eqref{eq:rx-1}, there exists an equivalent transmission and estimation
  strategy of the form:
  \begin{align}\label{eq:U_E}
    U_t &= \tilde f_t(E_t, S_{0:t-1}, Y_{0:t-1}), \\
    \hat X_t &= \tilde g_t(S_{0:t}, Y_{0:t}) + Z_t.
    \label{eq:hat_E}
  \end{align}
  Moreover, for any transmission and estimation strategies of the
  form~\eqref{eq:U_E}--\eqref{eq:hat_E}, there exist transmission and
  estimation strategies of the form~\eqref{eq:tx-2} and~\eqref{eq:rx-1} that
  are equivalent. 
\end{lemma}

The proof is given in Appendix~\ref{app:proof_lemma_U_hat_E}.

An implication of Lemma~\ref{lemma:U_hat_E} is that we may assume that the transmitter transmits $E_t$ and the receiver estimates
\[
  \hat E_t = \hat X_t - Z_t = \tilde g_t(S_{0:t},Y_{0:t}).
\]

For this model, we can further simplify the structures of optimal transmitter
and estimator as follows. 
\begin{theorem}\label{thm:SQC}
  In Problem~\ref{prob:finite} with first-order autoregressive source, we have
  that:
  \begin{enumerate}
    \item \emph{Structure of optimal estimation strategy:} At each time $t$,
      there is no loss of optimality in choosing the estimates $\{\hat E_t\}_{t \ge
      0}$ as
      \[
        \hat E_t = 0,
      \]
      or, equivalently, choosing the estimates $\{\hat X_t\}_{t \ge 0}$ as:
      $\hat X_0 = 0$, and for $t > 0$, 
      \begin{equation}
        \hat X_{t} = \begin{cases}
          a \hat X_{t-1}, & \mbox{if $Y_t = \BLANK$} \\
          Y_t, & \mbox{if $Y_t \in \reals$}
        \end{cases}
        \label{eq:rx-a}
      \end{equation}

    \item \emph{Structure of optimal transmission strategy:} There is no
      loss of optimality in restricting attention to transmission strategies
      of the form
      \begin{equation}\label{eq:u-opt-ar}
        U_t = \tilde f_t(E_t, S_{t-1}).
      \end{equation}

    \item \emph{Dynamic programming decomposition:} Recursively define the
      following value functions: for any $e \in \reals$ and $s \in
      \ALPHABET S$,
      \begin{align}
        J_{T+1}(e,s) &= 0, \label{eq:dp-J0}\\
        \intertext{and for $t \in \{T, \dots, 0\}$,} \label{eq:dp-J1}
        J_t(e,s) &= \min_{u \in \ALPHABET U} \bar H_t(e,s,u),
      \end{align}
      where 
      \begin{multline*}
        \bar H_t(e,s,u) = \lambda(u) + \sum_{s' \in \ALPHABET S} Q_{ss'} p(s', u) d(e) \\
        + \EXP[J_{t+1}(E_{t+1}, S_{t})\,|\, E_t = e, S_{t-1} = s, U_t = u].
      \end{multline*}

      Let $\tilde f^*_t(e,s)$ denote the arg min of the right hand side
      of~\eqref{eq:dp-J1}. Then the transmission strategy $\bm{\tilde f^*}
      = (\tilde f^*_0, \dots, \tilde f^*_T)$ is optimal.
  \end{enumerate}
\end{theorem}
See Appendix~\ref{app:SQC} for the proof.

\subsection{Monotonicity and quasi-convexity of the optimal solution}
For autoregressive sources we can establish monotonicity and quasi-convexity
of the optimal solution. To that end, let us assume the following.
\begin{assumption}\label{assump:sto-mono}
  The channel transition matrix $Q$ is stochastic monotone, i.e., for all $i,j
  \in \{1, \dots, n\}$ such that $i > j$ and for any $\ell \in \{0, \dots,
  n-1\}$,
  \[
    \sum_{k = \ell + 1}^n Q_{ik} \ge \sum_{k = \ell +1}^n Q_{jk}.
  \]
\end{assumption}

\begin{theorem}\label{thm:monotonic-threshold-fin}
  For any $t \in \{0,\dots, T\}$, we
  have the following:
  \begin{enumerate}
    \item For all $s \in \ALPHABET S$, $J_t(e,s)$ is even and quasi-convex in
      $e$.
  \end{enumerate}
  Furthermore, under Assumption~\ref{assump:sto-mono},
  \begin{enumerate}
    \setcounter{enumi}{1}
    \item For every $e \in \ALPHABET X$, $J_t(e,s)$ is decreasing in $s$. 
    \item For every $s \in \ALPHABET S$, the transmission strategy $\tilde
      f_t(e,s)$ is even and quasi-convex in~$e$. 
  \end{enumerate}
\end{theorem}

Sufficient conditions under which the value function and the optimal strategy
are even and quasi-convex are identified
in~\cite[Theorem~1]{JC-AM-mono-TAC2019}. Properties~1 and~3 follow because the
above model satisfies these sufficient conditions. Property~2 follows from
standard stochastic monotonicity arguments. The details are presented in the
supplementary material.

An immediate consequence of Theorem~\ref{thm:monotonic-threshold-fin} is the
following:
\begin{corollary}\label{cor:monotonic-threshold-fin}
  Suppose that Assumption~\ref{assump:sto-mono} is satisfied and $\ALPHABET U$
  is finite set given by $\ALPHABET U = \{0,u^{(1)}, \dots, u^{(m)}\}$. For
  any $i \in \{0,1,\dots,m\}$, define\footnote{%
    Note that $k^{(0)}_t(s) = 0$ and Theorem~\ref{thm:monotonic-threshold-fin}
  implies $k^{(i)}_t(s) \le k^{(i+1)}_t(s)$ for any $i \in \{0,1,\dots,m\}$.}
  \[
    k^{(i)}_t(s) \DEFINED \inf \{e \in \reals_{\ge 0} : \tilde f_t(e,s) = u^{(i)}\}.
  \]
  For ease of notation, define $k^{(m+1)}_t(s) = \infty$. 

  Then, the optimal strategy is a threshold based strategy given as follows:
  for any $s \in \ALPHABET S$, $i \in \{0, \dots, m\}$ and $|e| \in
  \big[k^{(i)}_t(s), k^{(i+1)}_t(s)\big)$, 
  \begin{equation}\label{eq:tilde-f-threshold}
    \tilde f_t(e,s) = u^{(i)}.
  \end{equation} 
\end{corollary}

\subsubsection*{Some remarks}
\begin{enumerate}
  \item It can be shown that under the optimal strategy, $\Pi^2_t$ is
    symmetric and unimodal ($\ASU$) (see Definition~\ref{def:SU}) around $\hat
    X_t$ and, therefore, $\Pi^1_t$ is $\ASU$ around $a \hat X_{t-1}$. Thus,
    the transmission and estimation strategies in Theorem~\ref{thm:SQC} depend
    on the pre- and post-transmission beliefs only through their means.

  \item Since the distortion function is even and quasi-convex, we can write
    the threshold conditions
    \[
      k^{(i-1)}_t(s) \le |e| < k^{(i)}_t(s)
    \]
    in~\eqref{eq:tilde-f-threshold} as
    \[
      d(k^{(i-1)}_t(s)) \le d(e) < d(k^{(i)}_t(s)).
    \]
    Thus, if we define distortion levels $D^{(i)}_t(s) =
    \big[d(k^{(i-1)}_t(s)), d(k^{(i)}_t(s))\big)$, then we can say that the
    optimal strategy is to transmit at power level $u_{(i)}$ if $E_t \in
    D^{(i)}_t(S_{t-1})$.

  \item When $Y_t = \BLANK$, the update of the optimal estimate is same as the
    update equation of Kalman filter. For this reason, we refer to the
    estimation strategy~\eqref{eq:rx-a} as a \emph{Kalman-filter like estimator}. 
\end{enumerate}

\subsection{Generalization to infinite horizon model}
\label{sec:infin_AR_results}

Given a communication strategy $(\VEC f, \VEC g)$, let $D_\beta^{(f, g)}(e,s)$
and $P_\beta^{(f, g)}(e,s)$ denote respectively the expected distortion and
expected transmitted prower when the system starts in state $(e,s)$, i.e., for
$\beta \in (0,1)$,
\begin{align*}
  D_\beta^{(f, g)}(e,s) &\DEFINED (1-\beta) \EXP^{(f,g)} \Big[\sum_{t = 0}^\infty \beta^t d(E_t)\,|\, E_0 = e, S_{-1} = s\Big],\\
  P_\beta^{(f, g)}(e,s) &\DEFINED (1-\beta) \EXP^{(f,g)} \Big[\sum_{t = 0}^\infty \beta^t \lambda(U_t)\,|\, E_0 = e, S_{-1} = s\Big],
\end{align*} 
and for $\beta = 1$,
\begin{align*}
  D_1^{(f, g)}(e,s) &\DEFINED \lim_{T \to \infty} \frac 1T \EXP^{(f,g)}
  \Big[\sum_{t = 0}^{T-1} d(E_t)\,|\, E_0 = e, S_0 = s \Big],\\
  P_1^{(f, g)}(e,s) &\DEFINED \lim_{T \to \infty} \frac 1T \EXP^{(f,g)}
  \Big[\sum_{t = 0}^{T-1} \lambda(U_t)\,|\, E_0 = e, S_0 = s \Big].
\end{align*}
Then, the performance of the strategy $(\VEC f, \VEC g)$ when the system starts in state $(e,s)$ is given by 
\[
  J_\beta^{(f,g)}(e,s) \DEFINED D_\beta^{(f, g)}(e,s) + P_\beta^{(f, g)}(e,s).
\]

The structure of optimal estimator, as established in Theorem~\ref{thm:SQC},
continues to hold for the infinite horizon setup as well. Thus, we can
restrict attention to Kalman-filter like estimator given by~\eqref{eq:rx-a} and look
at the problem of finding the best response transmission strategy. This is a
single agent stochastic control problem. If the per-step distortion is
unbounded, then we need the following assumption---which implies that there
exists a strategy whose performance is bounded---for the infinite horizon
problem to be meaningful. 

\begin{assumption}\label{assump:bounded-dist}
  Let $f^{(0)}$ denote the transmission strategy that always transmits at
  power level $u_{\max}$ and $g^*$ denote the Kalman-filter like strategy given
  by~\eqref{eq:rx-a}. Then, for given $\beta \in (0,1]$, and for all $e \in
  \ALPHABET X$ and $s \in \ALPHABET S$, $D^{(f^{(0)}, g)}_\beta(e,s) <
  \infty$.
\end{assumption}

Assumption~\ref{assump:bounded-dist} is always satisfied if $d(\cdot)$ is
bounded. For $\beta = 1$, $d(e) = e^2$ and $\ALPHABET S = \{0,1\}$, the
condition $a^2(1-Q_{00}) < 1$ is sufficient for
Assumption~\ref{assump:bounded-dist} to hold
(see~\cite[Theorem~8]{Huang_Dey_intermittentKF}
and~\cite[Corollary~12]{Rohretal_intermittentKF}). Similar sufficient
conditions are given in~\cite[Theorem~1]{wuetal2018} for vector-valued Markov
source processes with a Markovian packet-drop channel.

We now state the main theorem of this section.
\begin{theorem}\label{thm:SQC-infin}
In Problem~\ref{prob:infinite} with first-order autoregressive processes under
Assumption~\ref{assump:bounded-dist}, we have that
\begin{enumerate}
  \item \emph{Structure of optimal estimation strategy:} The time-homogeneous
    strategy $\bm{\tilde g^*} = \{\tilde g^*, \tilde g^*,\dots\}$, where
    $\tilde g^*$ is given by~\eqref{eq:rx-a}, is optimal.

  \item \emph{Structure of optimal transmission strategy:} There is no loss of
    optimality in restricting attention to time-homogeneous transmission
    strategies of the form
    \[
      U_t = \tilde f_\beta(E_t, S_{t-1}).
    \]

  \item \emph{Dynamic programming decomposition:} For $\beta \in (0,1)$, let
    $J_\beta$ be the smallest bounded solution of the following fixed point
    equation: for all $e \in \reals$ and $s \in \ALPHABET S$,
    \begin{equation}\label{eq:dp-J1-infin}
      J_\beta(e,s) = \min_{u \in \ALPHABET U} \bar H_\beta(e,s,u),
    \end{equation}
    where 
    \begin{multline*}
      \bar H_\beta(e,s,u) = (1-\beta)\lambda(u) + \sum_{s' \in \ALPHABET S} Q_{ss'} p(s', u) d(e) \\
      + \beta \EXP[J_\beta(E_{t+1}, S_{t})\,|\, E_t = e, S_{t-1} = s, U_t = u].
    \end{multline*}
    Let $\tilde f^*_\beta(e,s)$ denote the arg min of the right hand side
    of~\eqref{eq:dp-J1-infin}. Then the transmission $\bm{\tilde f^*_\beta} =
    (\tilde f^*_\beta, \tilde f^*_\beta, \dots)$ is optimal. \item
    \emph{Results for $\beta = 1$:} Let $\tilde f^*_1$ be any limit point of
    $\{\tilde f^*_\beta\}_{\beta \in (0,1)}$ as $\beta \uparrow 1$. Then,
    $\tilde f^*_1$ is optimal strategy for Problem~\ref{prob:infinite} with
    $\beta = 1$.
  \end{enumerate}
\end{theorem}

The proof is given in Appendix~\ref{app:AR-infin-proof}.

\begin{remark}\label{rem:non-uniqueness}
  We are not asserting that the dynamic program~\eqref{eq:dp-J1-infin} has a
  \emph{unique} fixed point. To make such an assertion, we would need to check
  the sufficient conditions for Banach fixed point theorem. These
  conditions~\cite{LermaLasserre} are harder to check than the sufficient
  conditions~(P1)--(P3) of Proposition~\ref{prop:p1-p3} that we verify in
  Appendix~\ref{app:AR-infin-proof}.
\end{remark}

\begin{corollary}\label{cor:monotonic-infin}
  The monotonicity properties of Theorem~\ref{thm:monotonic-threshold-fin}
  hold for the infinite horizon value function $J_\beta$ and transmission
  strategy $\tilde f_\beta$ as well.
\end{corollary}

An immediate consequence of Corollary~\ref{cor:monotonic-infin} is the following:
\begin{corollary}\label{cor:monotonic-threshold-infin}
  Suppose that Assumption~\ref{assump:sto-mono} is satisfied and $\ALPHABET U$
  is finite set given by $\ALPHABET U = \{0,u^{(1)}, \dots, u^{(m)}\}$. For
  any $i \in \{0,1,\dots,m\}$, define\footnote{%
    Note that $k^{(0)}_\beta(s) = 0$ and Corollary~\ref{cor:monotonic-infin}
  implies $k^{(i)}_\beta(s) \le k^{(i+1)}_\beta(s)$ for any $i \in
  \{0,1,\dots,m\}$.}
  \[
    k^{(i)}_\beta(s) \DEFINED \inf \{e \in \reals_{\ge 0} : \tilde f_(e,s) = u^{(i)}\}.
  \]
  For ease of notation, define $k^{(m+1)}_\beta(s) = \infty$. 

  Then, the optimal strategy is a threshold based strategy given as follows:
  for any $s \in \ALPHABET S$, $i \in \{0, \dots, m\}$ and $|e| \in
  \big[k^{(i)}_\beta(s), k^{(i+1)}_\beta(s)\big)$,
  \begin{equation}\label{eq:tilde-f-threshold-infin}
    \tilde f_\beta(e,s) = u^{(i)}.
  \end{equation} 
\end{corollary}

\section{Computing  optimal thresholds for autoregressive sources with finite actions}
\label{sec:finite_actions}
Suppose the power levels are finite and given by 
\[
  \ALPHABET U = \{0, u^{(1)}, \dots, u^{(m)}\}, \quad m \in \integers_{> 0},
\]
with $u^{(i)} < u^{(i + 1)}$ and $i \in \{0,1, \dots, m-1\}$. From
Corollary~\ref{cor:monotonic-threshold-infin}, we know that the optimal
strategy for Problem~\ref{prob:infinite} is a time-homogeneous threshold-based
strategy of the form~\eqref{eq:tilde-f-threshold}. Let $\mathbf k$ denote the
thresholds $\{k^{(i)}_\beta(s)\}$ and $f^{(\mathbf k)}$ denote the
strategy~\eqref{eq:tilde-f-threshold-infin}. In this section, we first derive
formulas for computing the performance of a general threshold-based strategy
$f^{(\mathbf k)}_\beta$ of the form~\eqref{eq:tilde-f-threshold} and then
propose a stochastic approximation based algorithm to identify the optimal
thresholds.

\begin{figure}[!t]
  \centering
  \includegraphics[width=\linewidth]{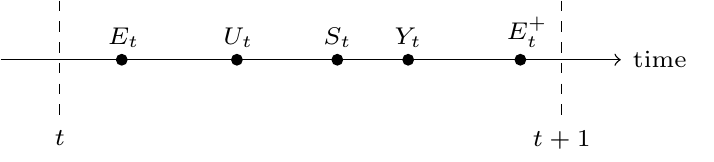}
  \caption{Timeline for pre- and post-transmission error state.}
  \label{fig:timeline}
\end{figure}

It is conceptually simpler to work with a \emph{post-decision model} where the
\emph{pre-decision state} is $E_t$ and the \emph{post-decision state} is
$E^+_t$ given by~\eqref{eq:e-update}. The timeline of the various system
variables is shown in Fig.~\ref{fig:timeline}. In this model, the per-step
cost is given by $\lambda(U_t) + d(E^+_t)$.\footnote{From
Theorem~\ref{thm:SQC}, we have that $\hat E_t = 0$. Thus, $d(E^+_t - \hat E_t)
= d(E^+_t)$.}

\subsection{Performance of an arbitrary threshold-based strategy}
For $\beta \in (0,1]$, pick a reference channel state $s^\circ \in \ALPHABET
S$. Given an arbitrary threshold-based strategy $f^{(\mathbf k)}_\beta$,
suppose the system starts in state $(E^+_{-1}, S_{-1}) = (0, s^\circ)$ and
follows strategy $f^{(\mathbf k)}_\beta$. Then, the process $\{(E^+_t,
S_t)\}_{t \ge 0}$ is a Markov process. Let $\tau^{(0)} = 0$ and for $n \in
\integers_{> 0}$ let
\[
  \tau^{(n)} \DEFINED \{ t > \tau^{(n-1)} : 
  (E^+_{t-1}, S_{t-1}) = (0, s^\circ) \}
\]
denote the \emph{stopping times} when the Markov process $\{(E^+_t, S_t)\}_{t
\ge 0}$ revisits $(0, s^\circ)$. We say that the Markov process
\emph{regenerates} at times $\{\tau^{(n)}\}_{n \in \integers_{\ge 0}}$ and
refer to the interval $\{\tau^{(n)}, \dots, \tau^{(n-1)} \}$ as the $n$-th
regenerative cycle. 

Define the following:
\begin{itemize}
 \item $L^{(\mathbf k)}_\beta$: the expected cost during a regenerative cycle,
   i.e., 
   \begin{multline}
     L^{(\mathbf k)}_\beta \DEFINED \EXP \Big[\sum_{t=0}^{\tau^{(1)}-1}\!\!
     \beta^t (\lambda(U_t) + d(E^+_t)) \Bigm| 
   \\ E^+_{-1}=0, S_{{-1}} = s^\circ \Big]. \label{eq:L-tau}
   \end{multline}

 \item $M^{(\mathbf k)}_\beta$: the expected time during a regenerative cycle,
   i.e., 
   \begin{equation}
     M^{(\mathbf k)}_\beta \DEFINED \EXP \Big[\sum_{t=0}^{\tau^{(1)}-1} \beta^t \,\Bigm|\, E^+_{-1}=0, S_{-1} = s^\circ \Big]. \label{eq:M-tau}
   \end{equation}
\end{itemize}
 
Using ideas from renewal theory, we have the following.
\begin{theorem}\label{thm:DNC}
 For any $\beta \in (0,1]$, the performance of threshold-based strategy
 $f^{(\mathbf k)}_\beta$ is given by
 \begin{equation}\label{eq:Ck}
   C^{(\mathbf k)}_\beta \DEFINED C_\beta(f^{(\mathbf k)}_\beta, g^*) = 
   \frac{ L^{(\mathbf k)}_\beta}{ M^{(\mathbf k)}_\beta}.
 \end{equation}
\end{theorem}
See Appendix~\ref{app:proof-thm_DNC} for the proof.

\subsection{Necessary condition for optimality}
In order to find the optimal threshold, we first observe the following.
\begin{lemma}\label{lemma:differentiable}
  For any $\beta \in (0,1]$, $L^{(\mathbf k)}_\beta$ and $M^{(\mathbf
  k)}_\beta$ are differentiable with respect to $\mathbf k$. Consequently,
  $C^{(\mathbf k)}_\beta$ is also differentiable.
\end{lemma}
The proof of Lemma~\ref{lemma:differentiable} follows from first principles
using an argument similar to that in the supplementary material for~\cite{JC_AM_TAC17}. 

Let $\nabla_{\mathbf k} L^{(\mathbf k)}_\beta$, $\nabla_{\mathbf k}
M^{(\mathbf k)}_\beta$ and $\nabla_{\mathbf k} C^{(\mathbf k)}_\beta$ denote
the derivatives of $L^{(\mathbf k)}_\beta$, $M^{(\mathbf k)}_\beta$ and
$C^{(\mathbf k)}_\beta$ respectively. Then, a sufficient condition for
optimality is the following.
\begin{proposition}\label{prop:necessary}
  A necessary condition for thresholds $\mathbf k^*$ to be optimal is that
  $N^{(\mathbf k^*)}_\beta = 0$, where
  \[
    N^{(\mathbf k^*)}_\beta \DEFINED M^{(\mathbf k)}_\beta \nabla_{\mathbf k} L^{(\mathbf k)}_\beta - L^{(\mathbf k)}_\beta \nabla_{\mathbf k} M^{(\mathbf k)}_\beta. 
  \]
\end{proposition}

\begin{proof}
  The result follows from observing that $\nabla_{\mathbf k} C^{(\mathbf
  k)}_\beta  = N^{(\mathbf k)}_\beta/(M^{(\mathbf k)}_\beta)^2$.
\end{proof}

\begin{remark}\label{rem:convex-C}
 If $C^{(\mathbf k)}_\beta$ is convex in $\mathbf k$, then the condition in
 Proposition~\ref{prop:necessary} is also sufficient for optimality. Based on
 numerical calculations we have observed that $C^{(\mathbf k)}_\beta$ is
 convex in $\mathbf k$ but we have not been able to prove it analytically.  
\end{remark}

\subsection{Stochastic approximation algorithm to compute optimal thresholds}
In this section we present an iterative algorithm based on simultaneous
perturbation and renewal Monte Carlo
(RMC)~\cite{JC-JS-AM-ACC17,JS-AM-RMC-arxiv} to compute the optimal thresholds.
We present this algorithm under the following assumption.
\begin{assumption}\label{assump:power-saturation}
  There exists a $K \in \ALPHABET X_{\ge 0}$ such that for the optimal
  transmission strategy
  \[
    k^{(m-1)}(s) \le K, \quad \forall s \in \ALPHABET S.
  \]
\end{assumption}

\begin{remark}\label{rem:power-saturation}
  Assumption~\ref{assump:power-saturation} is equivalent to stating that for
  each channel state $s$, there is a state $e^*(s)$ such that for all $e \ge
  e^*(s)$, the optimal transmission strategy transmits at the maximum power
  level $u_{\max}$ in state $(e,s)$. Assumption~\ref{assump:power-saturation}
  is similar to the channel saturation assumption
  in~\cite[Assumption~1]{Gatsisetal2014} and~\cite[Remark~5]{Renetal2017}. 
\end{remark}

Under Assumption~\ref{assump:power-saturation}, there is no loss of optimality
in restricting attention to threshold strategies in the set
 \[
   \mathcal K \DEFINED \{\mathbf k : k^{(i)}(s) \le K, \, \forall s \in \ALPHABET S, i \in \{0, \dots, m-1\}\}.
 \]

The main idea behind RMC is as follows. Given a threshold $\mathbf k$, consider the following sample-path based unbiased estimators of $L^{(\mathbf k)}_\beta$ and $M^{(\mathbf k)}_\beta$:
\begin{align}
  \widehat \JL^{(\mathbf k)}_\beta &\DEFINED \frac 1N \sum_{n = 0}^{N-1} \sum_{t = \tau^{(n)}}^{\tau^{(n+1)}-1} \beta^{t-\tau^{(n)}} \Big[\lambda(U_t) + d(E^+_t)\Big], \label{eq:hat-L}\\
  \widehat \M^{(\mathbf k)}_\beta &\DEFINED \frac 1N \sum_{n = 0}^{N-1} \sum_{t = \tau^{(n)}}^{\tau^{(n+1)}-1} \beta^{t-\tau^{(n)}}, \label{eq:hat-M}
\end{align}
where $N$ is a large non-negative integer.

Then, \emph{simultaneous perturbation} based unbiased estimators of $\nabla_{\mathbf k} L^{(\mathbf k)}_\beta$ and $\nabla_{\mathbf k} M^{(\mathbf k)}_\beta$ are given by
\begin{align}
  \widehat \nabla \JL^{(\mathbf k)}_\beta & = \delta (\widehat \JL^{(\mathbf k + c \delta)}_\beta - \widehat \JL^{(\mathbf k -c \delta)}_\beta)/ 2c, \label{eq:SFSA-L}\\
  \widehat \nabla \M^{(\mathbf k)}_\beta & = \delta (\widehat \M^{(\mathbf k + c \delta)}_\beta - \widehat \M^{(\mathbf k -c \delta)}_\beta)/ 2c \label{eq:SFSA-M},
\end{align}
where $\delta$ is an appropriately chosen random variable having the same
dimension as $\mathbf k$ and $c$ is a small positive constant. Typically, all
components of $\delta$ are chosen independetly as either
$\mathrm{Rademacher}(\pm 1)$~\cite{spall1992multivariate,MaryakChin} or
$\mathrm{Normal}(0,1)$~\cite{KK1972,Bhatnagar2013}.

If the estimates $(\widehat \JL^{(\mathbf k)}_\beta, \widehat \M^{(\mathbf
k)}_\beta)$ and $(\widehat \nabla \JL^{(\mathbf k)}_\beta, \widehat \nabla
\M^{(\mathbf k)}_\beta)$ are generated from from independent sample paths,
then the unbiasedness and independence of these estimates imply that
\[
  \widehat \N^{(\mathbf k)}_\beta \DEFINED \widehat \M^{(\mathbf k)}_\beta  \widehat \nabla \JL^{(\mathbf k)}_\beta  - \widehat \JL^{(\mathbf k)}_\beta \widehat \nabla \M^{(\mathbf k)}_\beta
\]
is an unbiased estimator of $N^{(\mathbf k)}_\beta$.

Then, the RMC algorithm to compute the optimal threshold is as follows.
\begin{enumerate}
 \item Let $j = 0$ and pick any initial guess $\mathbf k_0$.
 \item For sufficiently large number of iterations do
 \begin{enumerate}
  \item Generate $\widehat \JL^{(\mathbf k_j)}_\beta$ and $\widehat \M^{(\mathbf k_j)}_\beta$ according to~\eqref{eq:hat-L}--\eqref{eq:hat-M}.
  \item Sample a random direction $\delta$.
  \item Generate $\widehat \nabla \JL^{(\mathbf k_j)}_\beta$ and $\widehat \nabla \M^{(\mathbf k_j)}_\beta$ according to~\eqref{eq:SFSA-L}--\eqref{eq:SFSA-M}.
  \item Compute: 
  \[
    \widehat \N^{(\mathbf k_j)}_\beta \DEFINED \widehat \M^{(\mathbf k_j)}_\beta  \widehat \nabla \JL^{(\mathbf k_j)}_\beta  - \widehat \JL^{(\mathbf k_j)}_\beta \widehat \nabla \M^{(\mathbf k_j)}_\beta.
  \]
  \item Update:
  \[
    \mathbf k_{j+1} = [\mathbf k_j - \alpha_j \widehat \N^{(\mathbf k_j)}_\beta]_{\mathcal K},
  \]
  where $[\cdot]_{\mathcal K}$ denotes projection on to the set $\mathcal K$ and $\{\alpha_j\}$ is a sequence of learning rates that satisfy $\sum_{j = 1}^\infty \alpha_j = \infty$ and $\sum_{j = 1}^\infty \alpha^2_j < \infty$.
  \item Set $j = j+1$.
 \end{enumerate}
\end{enumerate}

We assume the following (which is a standard assumption for stochastic
approximation algorithms, see e.g., \cite[Assumption~5.6]{Bhatnagar2013}).
\begin{assumption}\label{assump:stable_eq}
 The set of globally asymptotically stable equilibrium of the ODE
 \begin{equation*}
   \frac{d\mathbf k}{dt} = - N{(\mathbf k)}_\beta
 \end{equation*}
 is compact.
\end{assumption}

\begin{theorem}\label{thm:RMC-local-minima}
  Consider the sequence of iterates $\{\mathbf k_j\}_{j \ge 0}$ obtained by
  the RMC algorithm described above. Let $\mathbf k^*$ be any limit point of
  $\{\mathbf k_j\}_{j \ge 0}$. Then, under
  Assumptions~\ref{assump:power-saturation} and~\ref{assump:stable_eq},
  $N^{(\mathbf k^*)}_\beta = 0$ and therefore $\nabla_{\mathbf k} C^{(\mathbf
  k^*)}_\beta = 0$.
\end{theorem}
\begin{proof}
  The proof follows from~\cite[Corollary~1]{JS-AM-RMC-arxiv}.
\end{proof}

\begin{figure*}[!t]
  \begin{subfigure}[b]{0.45\textwidth}
    \centering
    \includegraphics[page = 1, width=\textwidth]{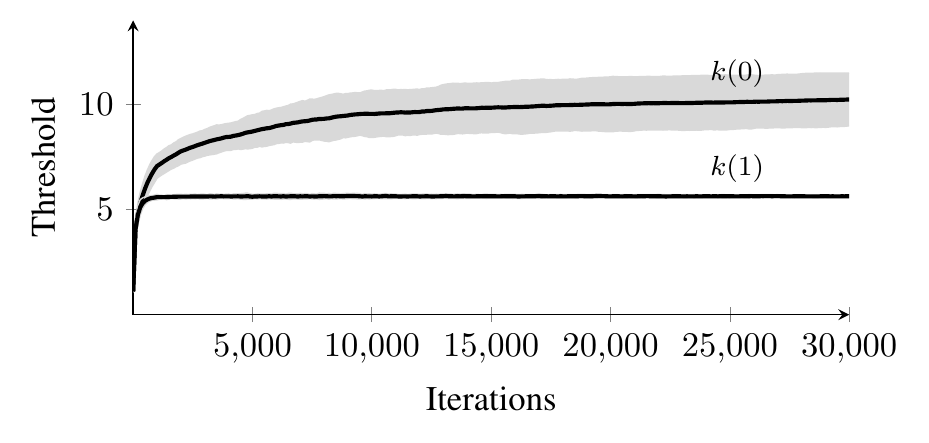}
    \caption{$\lambda_1 = 100$}   
    \label{fig:SF-optimal-k120}
  \end{subfigure}
  \hfill
  \begin{subfigure}[b]{0.45\textwidth}  
    \centering 
    \includegraphics[page = 2, width=\textwidth]{figures/k0_k1_vs_iterationsv5.pdf}
    \caption{$\lambda_1 = 200$}    
    \label{fig:SF-optimal-k560}
  \end{subfigure}
  \caption{The thresholds versus iterations for different values of $\lambda_1$. The experiment  is repeated 100 times. The bold lines represent the sample means and the shaded regions correspond to mean $\pm$ twice the standard deviation ($2\sigma$) across the~runs.} 
  \label{fig:sample-paths}
\end{figure*}

\subsection{Numerical example}
Consider a real-valued autoregressive source with $a = 1$, $\mu =
\mathrm{Normal}(0,1)$ and a Gilbert-Elliott channel~\cite{gilbert1960,
elliott1963} with state space
$\ALPHABET S = \{0,1\}$, transition matrix $Q = \begin{bsmallmatrix}
                                             0.3 & 0.7 \\
                                             0.1 & 0.9
                                           \end{bsmallmatrix}$,
power levels $\ALPHABET U = \{0,1\}$, loss probability 
\[
  p(0,0) = 1, \quad p(0,1) = 0.7, \quad p(1,0) = 1, \quad p(1,1) = 0.2,
\]
transmission cost $\lambda(0) = 0$, $\lambda(1) = \lambda_1$, and discount
factor $\beta = 0.9$. It can be verified that $Q$ is stochastic monotone.
Thus, Assumption~\ref{assump:sto-mono} is satisfied. 

We run the RMC algorithm with $s^\circ = 0$, $N = 1000$, the learning rates
$\{\alpha_j\}$ chosen according to ADAM~\cite{kingma_ba15} (with $\alpha$
parameter of ADAM equal to $0.1$ and other parameters taking their default values as stated in~\cite{kingma_ba15}), $\delta = \mathrm{Normal}(0,1)$ and $c = 0.1$. Note that since $m=1$, $k^{(0)}(s) = 0$, and we simply use $k(0)$ and $k(1)$ to denote $k^{(1)}(0)$ and $k^{(1)}(1)$. 

We pick 10 uniformly spaced values of $\lambda_1$ in $[50, 200]$ and run the
RMC algorithm for 30,000 iterations. Since the output is stochastic, we repeat
each experiment 100 times. We take the value of $\mathbf k$ at the end of each
run and compute $C^{(\mathbf k)}_\beta$ using $\widehat \C^{(\mathbf k)}_\beta
= \widehat \JL^{(\mathbf k)}_\beta / \widehat \M^{(\mathbf k)}_\beta$, where
we compute $\widehat \JL^{(\mathbf k)}_\beta$ and $\widehat \M^{(\mathbf
k)}_\beta$ by averaging over $N = 10^6$ renewals. The value of $\mathbf k$ at
the end of the run and $C^{(\mathbf k)}_\beta$ are shown in
Table~\ref{table:SF-GE-costly}. The table also shows the two-standard
deviation (denoted by $\sigma$) uncertainty on $\mathbf k$ and $C^{(\mathbf
k)}_\beta$.

A plot of $\mathbf k$ versus iterations for $\lambda_1 \in \{100, 200\}$ is shown in Fig.~\ref{fig:sample-paths}. This shows that the thresholds converge relatively quickly. 

\begin{table}[!b!t]
\centering
\caption{Optimal thresholds and performance for different values of
communication cost for $\beta = 0.9$}
\def\1{\multicolumn{1}{c}}
\begin{tabular}{@{}S[table-format=3.2]SSS@{}}
  \toprule
  \1{$\lambda_1$} & \1{$k(0)$} & \1{$k(1)$} & \1{$C^{(\mathbf k)}_{\beta}$} \\
  &\1{($\text{mean} \pm 2\sigma$)} & \1{($\text{mean} \pm 2\sigma$)} & \1{($\text{mean} \pm 2\sigma$)} \\
  \midrule
  50 & 8.669 \pm 1.408 & 4.465 \pm 0.061 & 4.991 \pm 0.015 \\
  66.67 & 9.438 \pm 1.592 & 4.914 \pm 0.056 & 5.440 \pm 0.009 \\
  83.33 & 9.843 \pm 1.212 & 5.290 \pm 0.061 & 5.791 \pm 0.008 \\
  100 & 10.235 \pm 1.259 & 5.635 \pm 0.063 & 6.087 \pm 0.009 \\
  116.67 & 10.377 \pm 1.227 & 5.943 \pm 0.063 & 6.334 \pm 0.013 \\
  133.33 & 10.821 \pm 1.053 & 6.226 \pm 0.068 & 6.557 \pm 0.010 \\
  150 & 10.991 \pm 1.203 & 6.492 \pm 0.065 & 6.750 \pm 0.017 \\
  166.67 & 11.238 \pm 1.105 & 6.744 \pm 0.076 & 6.911 \pm 0.008 \\
  183.33 & 11.523 \pm 1.089 & 6.975 \pm 0.080 & 7.070 \pm 0.007 \\
  200 & 11.660 \pm 1.056 & 7.203 \pm 0.086 & 7.198 \pm 0.014 \\
  \bottomrule
\end{tabular}
\label{table:SF-GE-costly}
\end{table}

\section{Discussions}
\subsection{Comparison with the results of~\cite{Renetal2017}}\label{sec:compare-ren}

Remote estimation over a packet-drop channel with Markovian state was recently considered in~\cite{Renetal2017}. In~\cite{Renetal2017} it is assumed that the transmitter knows the current channel state. In contrast, in our model we assume that the receiver observes the channel state and sends it back to the transmitter. So, the transmitter has access to a one-step delayed channel state.

In~\cite{Renetal2017}, the authors pose the problem of identifying the optimal
transmission and estimation strategies for infinite horizon average cost setup
for vector-valued autoregressive sources. They identify the common information
based dynamic program and identify technical conditions under which the
dynamic program has a deterministic solution. The dynamic program
in~\cite{Renetal2017} may be viewed as the infinite horizon average cost
equivalent of the finite horizon dynamic program in
Theorem~\ref{thm:structure-fin}. They then show that when the source dynamics
are orthogonal and the noise dynamics are isotropic, there is no loss of
optimality in restricting attention to estimation strategies of the
form~\eqref{eq:rx-*-fin} and transmission strategies of the
form~\eqref{eq:tx-*-fin}. In addition, for every $\pi \in \Delta(\ALPHABET X)$
and $s \in \ALPHABET S$, $\zeta_t(\cdot) \DEFINED f_t(\cdot, s, \pi)$ is
symmetric and quasi-convex. This structural property of the transmitter
implies that when the power levels are finite, there exist thresholds
$\{k^{i}_t(s,\pi)\}_{i=0}^{m-1}$ such that the optimal strategy is a threshold
based strategy as follows: for any $s \in \ALPHABET S$, $\pi \in
\Delta(\ALPHABET X)$, $i \in \{0, \dots, m\}$ and $|e| \in [k^{(i)}_t(s, \pi),
k^{(i+1)}_t(s, \pi))$,
\begin{equation}\label{eq:threshold-with-pi}
  f_t(e,s,\pi) = u^{(i)}.
\end{equation}

In this paper, we follow a different approach. We investigate both finite
Markov sources and first order autoregressive sources. For Markov sources, we
first show that there is no loss of optimality in restricting attention to
estimation strategies of the form~\eqref{eq:rx-*-fin} and transmission
strategies of the form~\eqref{eq:tx-*-fin}. For autoregressive sources, we
show that the structure of the transmission strategies can be further
simplified to~\eqref{eq:rx-a} and~\eqref{eq:u-opt-ar}. In addition for every
$s \in \ALPHABET S$, $\varphi_t(\cdot) = \tilde f_t(\cdot,s)$ is symmetric and
quasi-convex.  This structural property of the transmitter implies that when
the power levels are finite, there exist thresholds $\{k^i_t(s)\}_{i=0}^{m-1}$
auch that the optimal strategy is a threshold based strategy given by~\eqref{eq:tilde-f-threshold}. 

Once we restrict attention to estimation strategy of the form~\eqref{eq:rx-a}, the best response strategy at the transmitter is a centralized MDP. This allows us to establish the existence of optimal deterministic strategies for both discounted and average cost infinite horizon models without having to resort to the detailed technical argument presented in~\cite{Renetal2017}.

Note that in the threshold-based strategies~\eqref{eq:threshold-with-pi}
identified in~\cite{Renetal2017}, the thresholds depend on the belief state
$\pi$, while in the threshold-based strategies~\eqref{eq:tilde-f-threshold}
identified in this paper, the thresholds do not depend on $\pi$.  We exploit
this lack of dependence on $\pi$ to develop a renewal theory based method to
compute the performance of a threshold based strategy. The algorithm proposed
in this paper will not work for threshold strategies of the
form~\eqref{eq:threshold-with-pi} due to the dependence on $\pi$ (which is
uncountable). 

\subsection{Comparison with the results of~\cite{JC_AM_TAC17}}
A method for computing the optimal threshold for remote estimation over
noiseless communication channel (i.e., no packet drop) is presented
in~\cite{JC_AM_TAC17}. That method relies on computing $L^{(\mathbf k)}_\beta$
and $M^{(\mathbf k)}_\beta$ by solving the balance equations  (which are
Fredholm integral equations of the second kind) for the truncated Markov
chain. When the channel is a packet-drop channel, the kernel of the Fredholm integral
equation is discontinuous. Moreover, when the channel has state, the integral
equation is multi-dimensional. Solving such integral equations is
computationally difficult. The simulation based methods presented in this
paper circumvent these difficulties.

\subsection{The special case with i.i.d. packet-drop channels}\label{subsec:multiple_iid_channel}
Consider the case when the packet drops are i.i.d., which can be viewed as a
Markov channel with a single state (i.e., $|\ALPHABET S| = 1$). Thus we may
drop the dependence on $s$ from the value function and the strategies.
Furthermore, Assumption~\ref{assump:sto-mono} is trivially satisfied. Thus the
result of Theorem~\ref{thm:monotonic-threshold-fin} simplifies to the
following.
\begin{corollary}\label{cor:iid}
  For Problem~\ref{prob:finite} with i.i.d. packet drops the value function
  $J_t(e)$ and the optimal transmission strategy $f_t(e)$ are even and
  quasi-convex. 
\end{corollary}
The above result is same as~\cite[Theorem~1]{JC_AM_IFAC16}. Furthermore, when
the power levels are finite, the optimal transmission strategy is
characterized by thresholds $(k^{(1)}_t, \dots, k^{(m-1)}_t)$. For infinite
horizon models the thresholds are time-invariant. 

In addition, the renewal relationships of Theorem~\ref{thm:DNC} continues to
hold. The stopping times $\{\tau^{(n)}\}_{n \ge 0}$ correspond to times of
successful reception and the proposed renewal Monte-Carlo algorithm is similar
in spirit to~\cite{JC-JS-AM-ACC17}. Note that the algorithm proposed
in~\cite{JC-JS-AM-ACC17} uses simultaneous perturbation to find the minimum of
$C^{(k)}_\beta$, where $C^{(k)}_\beta$ is evaluated using renewal
relationships. The algorithm proposed in this paper is different and uses
simultaneous perturbations to find the roots of $N^{(\mathbf k)}_\beta$, which
coincide with the roots of $\nabla_{\mathbf k} C^{(\mathbf k)}_\beta$.

\begin{figure}[!t]
  \centering
  \includegraphics[width=\linewidth]{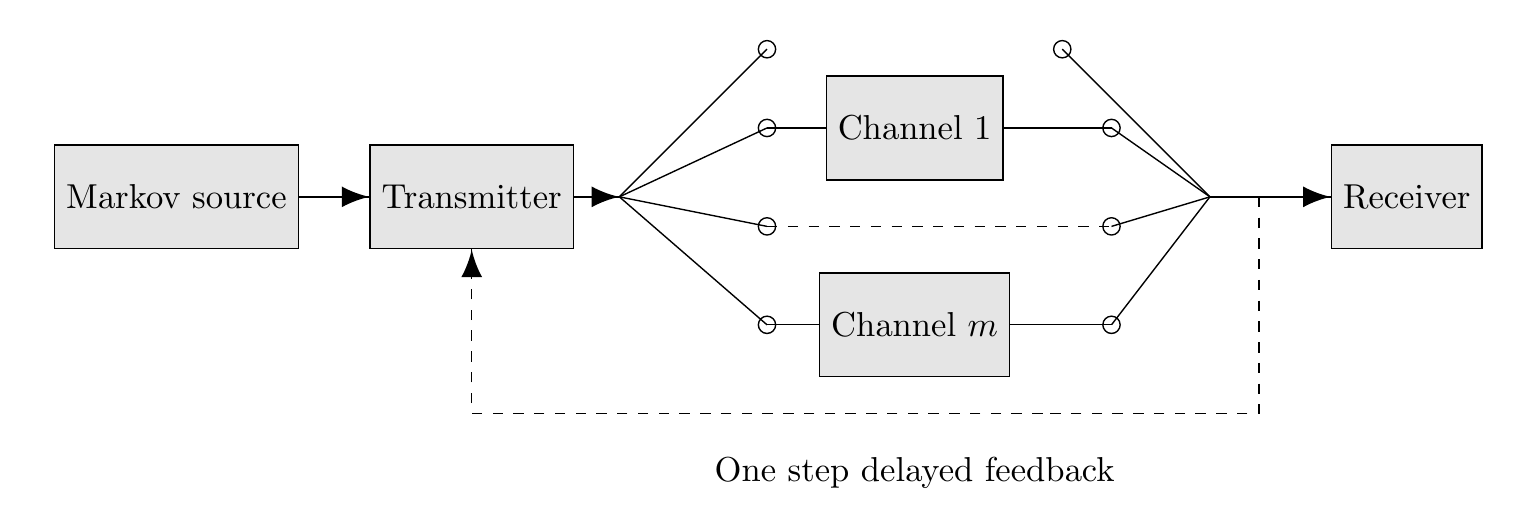}
  \caption{Remote estimation over multiple channels.}
  \label{fig:block_multiple_channels}
\end{figure}

It is worth highlighting that when the power levels are finite, the model can
also be interpreted as a remote estimator that has the option of transmitting
(at a constant power level) over one of $m$ available i.i.d.\@ packet-drop
channels, as shown in Fig~\ref{fig:block_multiple_channels}.  For channel $i$,
$i \in \{1, \dots, m\}$, the transmission cost is $\lambda(i)$ and the drop
probability is $p(i)$. We assume that the channels are ordered\footnote{% This
  assumption is without loss of generality. In particular, if there are
channels $i$ and $j$ such that $\lambda (i) \ge \lambda(j)$ and $p(i) \ge
p(j)$, then transmission over channel~$j$ dominates the action of transmission
over channel~$i$.} such that $\lambda(1) \le \dots \le \lambda(m)$ and $p(1)
\ge \dots \ge p(m)$. In addition, the sensor has the option of not
transmitting, which is denoted by $i = 0$. Note that $\lambda(0) = 0$ and
$p(0) = 1$. As argued above, the optimal transmission strategy in this case is
characterized by thresholds $(k^{(1)}_t, \dots, k^{(m-1)}_t)$ and the sensor
transmits over channel $i$, where $i$ is such that $|e_t| \in [k^{(i)}_t,
k^{(i+1)}_t)$ (and we assume that $k^{(0)}_t = 0$ and $k^{(m)}_t = \infty$).
The above result is similar in spirit to~\cite{Tameretalacc16}, which
considers i.i.d. source and additive noise channels.

\section{Conclusion}

In this paper, we study remote estimation over a Markovian channel with
feedback. We assume that the channel state is observed by the receiver and fed
back to the transmitter with one unit delay. In addition, the transmitter gets
\textsc{ack}/\textsc{nack} feedback for successful/unsuccessful transmission.
Using ideas from team theory, we establish the structure of optimal
transmission and estimation strategies for finite Markov sources and identify
a dynamic program to determine optimal strategies with that structure. We then
consider first-order autoregressive sources where the noise process has
unimodal and symmetric distribution. Using ideas from majorization theory, we
show that the optimal transmission strategy has a monotone structure and the
optimal estimation strategy is like Kalman filter. 

The structural results imply that threshold based transmitter is optimal when
the power levels are finite. We provide a stochastic approximation based
algorithm to compute the optimal thresholds and optimal performance. An
example of a first-order autoregressive source model with Gilbert-Elliott
channel is considered to illustrate the results.

\appendices
\section{Proof of Lemma~\ref{lemma:irrelevant_info_Tx}}
\label{app:proof-lemma-irr-info}

Arbitrarily fix the estimation strategy $\VEC g$ and consider the \emph{best
response} strategy at the transmitter. We will show that $\tilde I^1_t
\DEFINED (X_t, S_{0:t-1}, Y_{0:t-1})$ is an information state at the
transmitter. In particular, we will show that $\{\tilde I^1_t\}_{t \ge 1}$ satisfies the following properties:
\begin{align}
  \PR(\tilde I^1_{t+1}\,|\, I^1_t, U_t) &= \PR(\tilde I^1_{t+1}\,|\, \tilde I^1_t, U_t) \label{eq:equal_prob},
  \shortintertext{and}
  \EXP[c(X_t, U_t, \hat X_t)\,|\, I^1_t, U_t] &= \EXP[c(X_t, U_t, \hat X_t)\,|\, \tilde I^1_t, U_t].
  \label{eq:equal_exp}
\end{align}

Given any realization $(x_{0:T}, s_{0:T}, y_{0:T}, u_{0:T})$ of the system
variables $(X_{0:T}, S_{0:T}, Y_{0:T}, U_{0:T})$, define $i^1_t = (x_{0:t},
  s_{0:t-1},
y_{0:t-1}, \allowbreak u_{0:t-1})$ and $\tilde \imath^1_t = (x_t, s_{0:t-1},
y_{0:t-1})$. Now, for any $\breve \imath^1_{t+1} = (x_{t+1},
\breve s_{0:t}, \breve y_{0:t}) = (x_{t+1}, \breve s_t, \breve y_t,
\breve \imath^1_t)$, we use the shorthand $\PR(
\breve \imath^1_{t+1} | \tilde \imath^1_{0:t}, u_{0:t})$ to denote $\PR(\tilde
  I^1_{t+1} = \breve \imath^1_{t+1} | \tilde I^1_{0:t} = \tilde
i^1_{0:t}, U_{0:t} = u_{0:t})$. Then,
\begin{align}
  \hskip 2em & \hskip -2em
  \PR(\breve \imath^1_{t+1} | i^1_{t}, u_{t}) =
  \PR(x_{t+1}, \breve s_{t}, \breve y_{t}, \breve \imath^1_t | x_{0:t}, s_{0:t-1}, y_{0:t-1}, u_{0:t}) \notag \\
  &\stackrel{(a)}= 
  \PR(x_{t+1} | x_t) \PR(\breve y_t | x_t, \breve s_{t}) 
  \PR(\breve s_t | s_{t-1}) \notag \\
  & \hskip 2em \times \IND{\{ (\breve s_{0:t-1}, \breve y_{0:t-1}) = (s_{0:t-1}, y_{0:t-1})\}}
  \notag \\
  & = \PR(\breve \imath^1_{t+1} | x_t, s_{0:t-1}, y_{0:t-1}, u_t) \notag\\
  &=
  \PR(\breve \imath^1_{t+1} | \tilde \imath^1_t, u_t)
  \label{eq:Tr-Markov}
\end{align}
where $(a)$ follows from the source and the channel models. This shows that~\eqref{eq:equal_prob} is true.

Now consider~\eqref{eq:equal_exp}. Recall that $\hat X_t = g_t(I^2_t)$. Therefore, the expectation in the left hand side of~\eqref{eq:equal_exp} depends on $\PR(I^2_t\,|\, I^1_t,U_t)$. By marginalizing~\eqref{eq:Tr-Markov} with respect to $X_{t+1}$, we get
\[
  \PR(I^2_t\,|\, I^1_t,U_t) = \PR(I^2_t\,|\, \tilde I^1_t,U_t),
\]
which implies~\eqref{eq:equal_exp}.

Eq.~\eqref{eq:equal_prob} shows that $\{\tilde I^1_t\}_{t \ge 0}$ is a
controlled Markov process controlled by $\{U_t\}_{t \ge 0}$.
Eq.~\eqref{eq:equal_exp} shows that $\tilde I^1_t$ is sufficient for
performance evaluation. Hence, by Markov decision
theory~\cite{KumarVaraiya:1986}, there is no loss of optimality in
restricting attention to transmission strategies of the
form~\eqref{eq:tx-2}.

\section{Proof of Theorem~\ref{thm:structure-fin}}
\label{app:proof_thm-stru_fin}

Once we restrict attention to transmission strategies of the
form~\eqref{eq:tx-2}, the information structure is partial history
sharing~\cite{NMT:partial-history-sharing}. Thus, one can use the common
information approach of~\cite{NMT:partial-history-sharing} and obtain the
structure of optimal transmission strategy using this approach. 

Following~\cite{NMT:partial-history-sharing}, we split the information
available at each agent into a \emph{common information} and \emph{local
information}. Common information is the information available to all
decision makers in the future; the remaining data at the decision maker is
the local information. Thus, at the transmitter, the common information
is $C^1_t \DEFINED \{S_{0:t-1}, Y_{0:t-1}\}$ and the local information is
$L^1_t \DEFINED X_t$, and at the receiver, $C^2_t \DEFINED \{S_{0:t}, Y_{0:t}\}$ and $L^2_t = \emptyset$.   
The state sufficient for input output mapping of the system is $(X_t, S_t)$. 
By~\cite[Proposition 1]{NMT:partial-history-sharing}, we get that 
\begin{align*}
  \Theta^1_t(x,s) \DEFINED \PR(X_t = s, S_t = s | C^1_t)\\
  \Theta^2_t(x,s) \DEFINED \PR(X_t = s, S_t = s | C^2_t)
\end{align*}
are sufficient statistics for the common information at the transmitter and the receiver respectively. Now, we observe that
\begin{enumerate}
  \item $\Theta^1_t$ is equivalent to $(\Pi^1_t, S_{t-1})$ and $\Theta^2_t$ is equivalent to $(\Pi^2_t,S_t)$. This is because independence of $\{X_t\}_{t \ge 0}$ and $\{S_t\}_{t \ge 0}$ implies that $\theta^1_t(x,s) = \pi^1_t(x) Q_{s_{t-1}s}$ and $\theta^2_t(x,s) = \pi^2_t(x)\IND_{\{S_t=s\}}$.
  \item The expected distortion $D(\Pi^2_t, \hat X_t)$ does not depend on $S_t$ and the evolution of $\Pi^2_t$ to $\Pi^1_{t+1}$ (given by Lemma~\ref{lemma:F1-F2}) does not depend on $\hat X_t$.
\end{enumerate}
Thus, from~\cite[Proposition 1]{NMT:partial-history-sharing} we get that the optimal strategy is  given by the dynamic program of~\eqref{eq:dp-1-fin}--\eqref{eq:dp-2-fin}.

\section{Proof of Lemma~\ref{lemma:U_hat_E}}\label{app:proof_lemma_U_hat_E}
The proof relies on the fact that $Z_t$ is a deterministic function of $Y_{0:t}$, i.e., there exists an $\ell_t$ such that
\[
  Z_t = \ell_t (Y_{0:t}).
\]
We prove the two parts separately. We use the notation $\VEC f = (f_1, \dots, f_T)$, $\VEC g = (g_1, \dots, g_T)$, $\bm{\tilde f} = (\tilde f_1, \dots, \tilde f_T)$ and $\bm{\tilde g} = (\tilde g_1, \dots, \tilde g_T)$.
\begin{enumerate}
  \item Given a transmission and an estimation strategy $(\VEC f, \VEC g)$ of the form~\eqref{eq:tx-2} and~\eqref{eq:rx-1}, define
  \begin{multline*}
  \tilde f_t(E_t, S_{0:t-1}, Y_{0:t-1}) 
  \\ = f_t(E_t + a \ell_{t-1}(Y_{0:t-1}), S_{0:t-1}, Y_{0:t-1})
  \end{multline*}
  and
  \[
  \tilde g_t(S_{0:t}, Y_{0:t}) = g_t(S_{0:t}, Y_{0:t}) - \ell_t(Y_{0:t}).
  \]
  Then, by construction the strategy $(\bm{\tilde f}, \bm{\tilde g})$ is equivalent to $(\VEC f,\VEC g)$.
  \item Given a transmission and an estimation strategy $(\bm{\tilde f}, \bm{\tilde g})$ of the form~\eqref{eq:U_E}--\eqref{eq:hat_E}, define
  \begin{multline*}
  f_t(X_t, S_{0:t-1}, Y_{0:t-1}) \\
  = \tilde f_t(X_t - a \ell_{t-1}(Y_{0:t-1}), S_{0:t-1}, Y_{0:t-1})
  \end{multline*}
  and
  \[
  g_t(S_{0:t}, Y_{0:t}) = \tilde g_t(S_{0:t}, Y_{0:t}) + \ell_t(Y_{0:t}).
\]
  Then, by construction the strategy $(\VEC f,\VEC g)$ is equivalent to $(\bm{\tilde f}, \bm{\tilde g})$.
\end{enumerate}

\section{Proof of Theorem~\ref{thm:SQC}}\label{app:SQC}
\subsection{Sufficient statistic and dynamic program}\label{sec:suff-stat}

Similar to the construction of a prescription for the finite state Markov
sources, for any transmission strategy $\bm{\tilde g}$ of the
form~\eqref{eq:U_E} and any realization $(s_{0:t-1}, y_{0:t})$ of
$(S_{0:t-1},Y_{0:t})$, define $\varphi: \reals \to \ALPHABET U$ as:
\[
  \varphi(e) = \tilde f_t (e, s_{0:t-1}, y_{0:t-1}),\, \forall e \in \reals.
\]
Next, redefine the pre- and post-transmission beliefs in terms of the error
process. In particular, $\pi^1_t$ is the conditional pdf of $E_t$ given
$(s_{0:t-1}, y_{0:t-1})$ and $\pi^2_t$ is the conditional pdf of $E^+_t$ given
$(s_{0:t}, y_{0:t})$.

Let $R_t = \IND_{\{Y_t \in \ALPHABET X\}}$ and $r_t$ denote the realization of $R_t$. The time evolution of $\pi^1_t$ and $\pi^2_t$ is similar to
Lemma~\ref{lemma:F1-F2}. 
In particular, we have
\begin{lemma} \label{lemma:F1-F2a}
  Given any transmission strategy $\VEC f$ of the form~\eqref{eq:tx-1}:
  \begin{enumerate} 
    \item there exists a function $F^1$ such that
      \begin{equation} \label{eq:update-2a}
        \pi^1_{t+1} = F^1(\pi^2_t) = \tilde \pi^2_t \star \mu,
      \end{equation}
      where $\tilde \pi^2_t$ given by $ \tilde \pi^2_t(e) \DEFINED (1/|a|) \pi^2_t(e/a)$ is the conditional probability density of $aE^+_t$, $\mu$ is the probability density function of $W_t$ and $\star$ is the convolution operation.

    \item There exists a function $F^2$ such that for any realization $r_t \in \{0,1\}$ of $R_t$
      \begin{equation*} 
        \pi^2_t = F^2(\pi^1_t, s_{t}, \varphi_t, r_t).
      \end{equation*}
      In particular,
      \begin{equation}\label{eq:update-1a}
        \pi^2_t = F^2(\pi^1_t, s_{t}, \varphi_{t}, r_t)
        = \begin{cases}
        \pi^1_t|_{\varphi_t,s_{t}}
          , & \mbox{if $r_t = 0$}\\
          \delta_{0} & \mbox{if $r_t = 1$}.
        \end{cases}
      \end{equation}
  \end{enumerate}
\end{lemma}

The dynamic program of Theorem~\ref{thm:structure-fin} can be rewritten in terms of the error process as given below. Consider $\ALPHABET X = \reals$ (similar derivation holds for $\ALPHABET X = \integers$).
\begin{align}
  V^1_{T+1}(\pi^1_t,s_t) &= 0, \notag\\
  \intertext{and for $t \in \{T, \dots, 0\}$}
  V^1_t(\pi^1_t,s_t) &= \min_{\varphi_t \colon \ALPHABET X \to \ALPHABET U} 
  \Big\{ \Lambda (\pi^1_t, \varphi_t) +
    H_t(\pi^1_t, s_t, \varphi_t) 
  \Big\}
  \label{eq:dp-1-fin-e}
  \\
  V^2_t(\pi^2_t,s_t) &= D(\pi^2_t) 
  + V^1_{t+1}(\pi^2_t P,s_t),
  \label{eq:dp-2-fin-e}
\end{align}
where, 
\begin{align*}
  \Lambda (\pi^1_t, \varphi_t) &\DEFINED \int_{\reals} \lambda(\varphi(e)) \pi^1_t(e)de, \\
  H_t(\pi^1_t, s_t, \varphi) & \DEFINED
  B(\pi^1_t, s_t, \varphi_t) V^2_t(\delta_0,s_t)
  +\\
  & \hskip 2em (1-B(\pi^1_t, s_t, \varphi_t)) V^2_t(\pi^1|_{\varphi,s_t},s_t),
  \\
  D(\pi^2_t) & \DEFINED \min_{\hat e \in \ALPHABET X}
  \int_{\reals} d(e - \hat e) \pi^2_t(e) de.
\end{align*}

Note that due to the change of variables, the expression for the first term of
$H_t$ does not depend on the transmitted symbol. Consequently, the expression
for $V^1_t$ is simpler than that in Theorem~\ref{thm:structure-fin}.

\begin{remark}
  The common-information approach for decentralized stochastic control for
  finite state and finite action models was presented
  in~\cite{NMT:partial-history-sharing}. In general, to extend the approach to
  Borel state and action spaces, one needs to impose a topology on the space
  of prescriptions and establish an appropriate ``measurable selection
  theorem''. Such technical difficulties are not present in the dynamic
  program of \eqref{eq:dp-1-fin-e}--\eqref{eq:dp-2-fin-e} because the common
  information is finite\footnote{In particuar, $C^1_t = R_{0:t-1}$ and $C^2_t
  = R_{0:t}$.} and, therefore, all prescriptions are measurable. 
\end{remark}

To establish the results of Theorem~\ref{thm:SQC}, we show that the above dynamic program satisfies a monotonicity property with respect to the partial order based on majorization. We start with some mathematical preliminaries needed to present this argument.
\subsection{Mathematical preliminaries}
\begin{definition}[Symmetric and unimodal density]
  \label{def:SU}
  A probability density function $\pi$ on reals is said to be \emph{symmetric
  and unimodal ($\ASU$)} around $c \in \reals$ if for any $x \in \reals$,
  $\pi(c-x) = \pi(c+x)$ and $\pi$ is non-decreasing in the interval $(-\infty,
  c]$ and non-increasing in the interval $[c, \infty)$. 
\end{definition}

\begin{definition}[Symmetric and quasi-convex prescription]
  \label{def:SQC}
  Given $c \in \reals$, a prescription $\theta \colon \reals \to \ALPHABET U$
  is symmetric and quasi-convex (denoted by $\SQC(c)$) if $\theta(e-c)$ is
  even and quasi-convex. 
\end{definition}

Now, we state some properties of symmetric and unimodal distributions. 

\begin{property}\label{prop:1}
  If $\pi$ is $\ASU(c)$, then 
  \[
    c \in \arg \min_{\hat e \in \reals} \int_\reals
    d(e - \hat e) \pi(e) de.
  \]
\end{property}
\begin{proof}
  For $c=0$, the above property is a special case of~\cite[Lemma
  12]{LipsaMartins:2011}. The result for general $c$ follows from a change of
  variables.
\end{proof}

\begin{property}\label{prop:2}
  If $\pi^1_t$ is $\ASU(0)$ and $\varphi_t \in \SQC(0)$, then for any $r_t \in
  \{0,1\}$ and $s_t \in \ALPHABET S$, $F^2(\pi^1_t, s_t, \varphi_t, r_t)$ is
  $\ASU(0)$. 
\end{property}
\begin{proof}
  We prove the result for each $r_t \in \{0, 1\}$ separately. Recall the
  update of $\pi^1_t$ given by~\eqref{eq:update-1a}. 
  \begin{itemize}
    \item For $r_t = 0$ and a given $s_t \in \ALPHABET S$, we have that
   if $\varphi_t \in \SQC(0)$, then $p(s_t, \varphi_t(e)$ is $\ASU(0)$ since $p(s_t,\cdot)$ is decreasing and $p(s_t, \varphi_t(e)) = p(s_t, \varphi(-e))$. Then $\pi^1_t(e)p(s_t, \varphi_t(e))$ is $\ASU(0)$ since the product of two $\ASU(0)$ functions is $\ASU(0)$. Hence $\pi^2_t$ is $\ASU(0)$.   
   \item For $r_t = 1$, $\pi^2_t = \delta_0$, which is $\ASU(0)$. 
\end{itemize}
\end{proof}

\begin{property}\label{prop:3}
  If $\pi^2_t$ is $\ASU(0)$, then $F^1(\pi^2_t)$ is also $\ASU(0)$.
\end{property}
\begin{proof}
  Recall that $F^1$ is given by~\eqref{eq:update-2a}. The property follows
  from the fact that convolution of symmetric and unimodal distributions is
  symmetric and unimodal~\cite{LipsaMartins:2011}.
\end{proof}

\begin{definition}[Symmetric rearrangement (SR) of a set]
  \label{def:SR}
  Let $\ALPHABET A$ be a measurable set of finite Lebesgue measure, its
  \emph{symmetric rearrangement} $\ALPHABET A^\sigma$ is the open interval
  centered around origin whose Lebesgue measure is same as $\ALPHABET A$.
\end{definition}

\begin{definition}[Level sets of a function]
  \label{def:LS}
  Given a function $\ell \colon \reals \to \reals_{\ge 0}$, its upper-level
  set at level $\rho$, $\rho \in \reals$, is $\{ x \in \reals : \ell(x) > \rho
  \}$ and its lower-level set at level $\rho$ is $\{ x \in \reals : \ell(x) <
  \rho \}$.
\end{definition}

\begin{definition}[SR of a function]
  \label{def:SR-func}
  The \emph{symmetric decreasing rearrangement} $\ell^{\sigma}_\downarrow$ of
  $\ell$ is a symmetric and decreasing function whose level sets are the same
  as $\ell$, i.e., 
  \[
    \ell^{\sigma}_\downarrow(x) = \int_{\reals_{\ge 0}}
    \IND_{\{z \in \reals : \ell(z) > \rho\}^\sigma}(x) d\rho.
  \]
  Similarly, the \emph{symmetric increasing rearrangement}
  $\ell^{\sigma}_\uparrow$ of $\ell$ is a symmetric and increasing function
  whose level sets are the same as $\ell$, i.e., 
  \[
    \ell^{\sigma}_\uparrow(x) = \int_{\reals_{\ge 0}}
    \IND_{\{z \in \reals : \ell(z) < \rho\}^\sigma}(x) d\rho.
  \]
\end{definition}

\begin{definition}[Majorization]\label{def:majorization}
  Given two probability density functions $\xi$ and $\pi$ over $\reals$, $\xi$
  \emph{majorizes} $\pi$, which is denoted by $\xi \succeq_m \pi$, if for all
  $\rho \ge 0$, 
  \[
    \int_{|x| \ge \rho} \xi^\sigma_\downarrow(x) dx \ge 
    \int_{|x| \ge \rho} \pi^\sigma_\downarrow(x) dx .
  \]
\end{definition}

\begin{definition}[SU-majorization]\label{def:ASU-majorization}
  Given two probability density functions $\xi$ and $\pi$ over $\reals$, $\xi$
  \emph{SU majorizes} $\pi$, which we denote by $\xi \succeq_a \pi$,
  if $\xi$ is $\ASU$ and $\xi$ majorizes $\pi$.
\end{definition}

An immediate consequence of Definition~\ref{def:ASU-majorization} is the
following:
\begin{lemma}\label{lemma:SU-major-inequality}
  For any non-negative $\SQC(c)$, function $g$, $c \in \reals$, and given two
  probability density functions $\xi$ and $\pi$ over $\reals$, such that $\xi
  \succeq_a \pi$ and $\xi$ is $\ASU(c)$, we have that for any $A \subseteq
  \reals$,
  \[
    \int_A g(x) \xi(x) dx \ge \int_A g(x) \pi(x) dx.
  \]
%with equality for $A = \reals$.
\end{lemma}

\begin{property}\label{prop:4}
  For any $\xi \succeq_a \pi$, $F^1(\xi) \succeq_a F^1(\pi)$. 
\end{property}
This follows from~\cite[Lemma 10]{LipsaMartins:2011}.

Recall the definition of $D(\pi^2)$ given after~\eqref{eq:dp-2-fin-e}.

\begin{property}\label{prop:5}
  If $\xi \succeq_a \pi$, then 
  \[
     D(\pi) \ge D(\pi^\sigma_\downarrow) \ge D(\xi^\sigma_\downarrow) = D(\xi).
  \]
\end{property}
\begin{proof}
  The inequalities follow from~\cite[Lemma~11]{LipsaMartins:2011}. The last
  equality holds since $\xi$ is $\ASU(0)$ and thus $\xi^\sigma_\downarrow =
  \xi$.
\end{proof}

\begin{lemma}\label{lemma:pres}
  For any $c \in \reals$, densities $\pi$ and $\xi$ and a prescription
  $\varphi$, there exists a $\SQC(c)$ prescription $\theta$, $c \in \reals$,
  such that for any $u \in \ALPHABET U$
  \[
    \pi(\{e \in \reals\,:\, \varphi(e) \le u\}) = \xi(\{e \in \reals\,:\, \theta(e) \le u\}).
  \]
  We denote such a $\theta$ by $\mathcal T^{(\pi, \xi)}_c \varphi$.
\end{lemma}
\begin{proof}
  Let $\eta(u) \DEFINED \pi(\{e \in \reals\,:\, \varphi(e) \le u\})$. We prove
  the construction separately for $\ALPHABET X = \reals$ and $\ALPHABET X =
  \integers$. First, let us consider $\ALPHABET X = \reals$. Let $A(u)$ be a
  symmetric set centered around $c \in \reals$ such that $\xi(A(u)) =
  \eta(u)$. By construction, $\eta(u)$ is increasing in $u$ and therefore so
  is $A(u)$. Define $\theta(e)$ such that $A(u)$ is the lower level set of
  $\theta(e)$. By construction, $\theta(e)$ is symmetric around $c$. Moreover,
  since $A(u)$ is convex, $\theta(e)$ is quasi-convex.

  Now, let us consider $\ALPHABET X = \integers$ and $c=0$. (The proof for
  general $c$ follows from a change of variables). Let $\ALPHABET U_\varphi$
  denote the set $\{\varphi(e)\,:\, e \in \integers\}$. Note that $\ALPHABET
  U_\varphi$ is a finite or a countable set which we will index by
  $\{u^{(i)}\,;\, i \in \integers_{\ge 0}\}$. Define $k^{(i)}$ such that
  \[
    \xi(A^{(i-1)}) < \eta^{(i)} \le \xi(A^{(i)}),
  \]
  where $A^{(i-1)} \DEFINED \{-k^{(i-1)},\dots,k^{(i-1)}\}$ and $A^{(i)}
  \DEFINED \{-k^{(i)}, \dots, k^{(i)}\}$.

  Now, for any $u^{(i)} \in \ALPHABET U_\varphi$, let $\eta^{(i)}$ denote
  $\eta^{(i)} \DEFINED \pi(\{e \in \integers\,:\, \varphi(e) \le u^{(i)}\})$.
  Define $\delta^{(i)} \DEFINED (\eta^{(i)} - \xi(A^{(i)}))/2$. Define the
  prescription $\theta$ as follows
  \[
    \theta(e) = \begin{cases}
      u^{(i-1)}, & \mbox{if $|e| \in \big(k^{(i-1)}, k^{(i)} \big)$}\\
      u^{(i-1)} \text{ w.p. $\delta^{(i-1)}$}, & \mbox{if $|e| = k^{(i)}$}\\
      u^{(i)} \text{ w.p. $\delta^{(i)}$}, & \mbox{if $|e| = k^{(i)}$}.
    \end{cases}
  \]
  Then, by construction, $\theta$ is $\SQC(0)$ and $\xi(\{e \in \integers\,:\,
  \theta(e) \le u\}) = \eta(u)$. This completes the proof.
\end{proof}

\begin{remark}\label{rem:probU}
  Suppose $\theta = \mathcal T^{(\pi,\xi)}_c \varphi$. Then the probability of
  using power level $u$ at pre-transmission belief $\pi$ and prescription
  $\varphi$ is the same as that at pre-transmission belief $\pi$ and
  prescription $\theta$. In particular, for all $u \in \ALPHABET U$,
  \[
    \pi(\{e \in \ALPHABET X : \varphi(e) = u \}) =
    \xi(\{e \in \ALPHABET X : \theta (e) = u \}).
  \]
\end{remark}

\begin{property}\label{prop:6}
  For any density $\xi$, $\pi$,  and prescription $\varphi$ let $\theta = \mathcal T^{(\pi, \xi)}_c \varphi$. Then, for any $s \in \ALPHABET S$,
  \begin{enumerate}
  \item $B(\xi, s, \theta) = B(\pi, s, \varphi)$.
  \item $\Lambda(\pi, \varphi) = \Lambda(\xi,\theta)$.
  \end{enumerate}
\end{property}
The above results follow from the definitions of $B$ and $\Lambda$ and Remark~\ref{rem:probU}. See supplementary material for a detailed proof.

\begin{property} \label{prop:7}
For any $\xi \succeq_a \pi$, where $\xi$ is $\ASU(c)$, and prescription $\varphi$, let $\theta = \mathcal T^{(\xi, \pi)}_c \varphi$. Then, for any $s \in \ALPHABET S$
\begin{equation} \label{eq:7-1}
  \xi|_{\theta,s} \succeq_a \pi|_{\varphi,s}.
\end{equation}
Consequently, for any $s \in \ALPHABET S$ and $r \in \{0,1\}$, we have
\begin{equation} \label{eq:7-2}
    F^2(\xi, s, \theta, r) \succeq_a F^2(\pi, s, \varphi, r).
\end{equation}
\end{property}

\begin{proof}
We prove the result for finite $\ALPHABET U = \{u^{(0}, \dots, u^{(m)} \}$. The result for the case when $\ALPHABET U$ is an interval follows from a discretization argument.

For any $i \in \{0, \dots, m\}$, let $A^{(i)} = \{e\,:\, \theta(e) = u^{(i)} \}$ and $B^{(i)} = \{e : \varphi(e) = u^{(i)} \}$ and $\bar A^{(i)} = \bigcup_{j=0}^i A^{(i)}$ and $B^{(i)} = \bigcup_{j=0}^i B^{(i)}$. 

Since $\theta$ is $\SQC(0)$, $\bar A^{(i)}$ is an interval (while $\bar
B^{(i)}$ need not be an interval). Define $a^{(i)} = \xi(\bar A^{(i)})$ and
$b^{(i)} = \pi(\bar B^{(i)})$. Also define $\xi^{(i)}$ as $\xi^{(i)}(e) =
\xi(e) \IND_{\{e \in \bar A^{(i)}\}}/a^{(i)}$ and $\pi^{(i)} = \pi^{(i)}(e) =
\pi(e) \IND_{\{e \in \bar B^{(i)} \}}/b^{(i)}$. Then,
by\cite[Lemma~8]{LipsaMartins:2011},
\begin{equation}\label{eq:sd-relation}
  \xi^{(i)} \succeq_a \pi^{(i)}, \quad 
  \forall i \in \{0, \dots, m\}.
\end{equation}

Fix an $s \in \ALPHABET S$. For ease of notation, define $p^{(i)} = p(s, u^{(i)})$. Then, we can write the following expression for $\xi\big|_{\theta,s}$
\begin{align}
  \xi&\big|_{\theta,s}(e)\notag 
  = \frac{1}{B(\xi,s,\theta)} \xi(e)p(s, \theta(e)) \notag\\
  &= \frac{\xi(e)}{B(\xi,s,\theta)} \sum_{i=0}^m p^{(i)}\IND_{\{e \in A^{(i)}\}} \notag\\
  &\stackrel{(a)}= \frac{\xi(e)}{B(\xi,s,\theta)} \sum_{i=0}^{m-1} [p^{(i)}- p^{(i+1)}]\IND_{\{e \in \bar A^{(i)}\}}
  + p^{(m)}\IND_{\{e \in \bar A^{(m)}\}} \notag\\
  &= \frac{1}{B(\xi,s,\theta)} \sum_{i=0}^{m-1} a^{(i)} [p^{(i)} - p^{(i+1)}]
  \xi^{(i)}
  + a^{(m)} p^{(m)} \xi^{(m)},
\label{eq:xi_theta}
\end{align}
where $(a)$ uses the fact that $A^{(i)} = \bar A^{(i)} \setminus \bar
A^{(i-1)}$. By a similar argument, we have
\begin{multline}
  \pi\big|_{\varphi,s}(e) 
   = \frac{1}{B(\pi,s,\varphi)} \sum_{i=0}^{m-1} b^{(i)} [p^{(i)} -
  p^{(i+1)}]\pi^{(i)} \\
  + b^{(m)}p^{(m)} \pi^{(m)}.
\label{eq:pi_varphi}
\end{multline}
Property~\ref{prop:6} implies that $a^{(i)} = b^{(i)}$. The monotonicity of
$p(s,u)$ implies that $p^{(i)} - p^{(i+1)} \ge 0$. Using this, and
combining~\eqref{eq:xi_theta} and~\eqref{eq:pi_varphi}
with~\eqref{eq:sd-relation}, we get~\eqref{eq:7-1}.
Eq.~\eqref{eq:7-2} follows from~\eqref{eq:F2}.
\end{proof}

\subsection{Qualitative properties of the value function and optimal strategy}\label{sec:SQC-value-strategy}

\begin{lemma} \label{lemma:property}
  The value functions~$V^1_t$ and $V^2_t$
  of~\eqref{eq:dp-1-fin-e}--\eqref{eq:dp-2-fin-e}, satisfy the following property. 
  \begin{enumerate}
    \item [(V1)] For any $i \in \{1,2\}$, $s \in \ALPHABET S$, $t \in \{0,
      \dots, T\}$, and pdfs $\xi^i$ and $\pi^i$ such that $\xi^i \succeq_a
      \pi^i$, we have that $V^i_t(\xi^i,s) \le V^i_t(\pi^i,s)$. 
  \end{enumerate}

  Furthermore, the optimal strategy satisfies the following properties.  For
  any $s \in \ALPHABET S$ and $t \in \{0, \dots, T\}$:
  \begin{enumerate}
    \item[(V2)] if $\pi^1_t$ is $\ASU(c)$, then there exists a prescription
      $\varphi_t \in \SQC(c)$ that is optimal. In general, $\varphi_t$
      depends on $\pi^1_t$. 
    \item[(V3)] if $\pi^2_t$ is $\ASU(c)$, then the optimal estimate $\hat E_t$
      is $c$.
  \end{enumerate}
\end{lemma}

\begin{proof}
  We proceed by backward induction. $V^i_{T+1}(\pi^1,s)$, $i \in \{1,2\}$ trivially satisfy (V1). This forms the basis of induction. Now assume that $V^1_{t+1}(s,
  \pi^1)$ also satisfies (V1). For $\xi^2 \succeq_a \pi^2$, we have that
  \begin{align}
    V^2_t(\pi^2,s) &= D(\pi^2) + V^1_{t+1}(F^1(\pi^2),s) \notag \\
    &\stackrel{(a)}\ge D(\xi^2) + V^1_{t+1}(F^1(\xi^2),s) 
    = V^2_t(\xi^2,s),
    \label{eq:induction-2}
  \end{align}
  where $(a)$ follows from Properties~\ref{prop:4} and~\ref{prop:5} and the
  induction hypothesis. Eq.~\eqref{eq:induction-2} implies that $V^2_t$ also
  satisfies (V1). Now, we have
  \begin{align}
  &H_t(\pi^1,s,\varphi) = B(\pi^1, s, \varphi) V^2_t(\delta_0,s)
   + \notag\\ 
   & \hskip 8em (1-B(\pi^1, s, \varphi)) V^2_t(\pi^1|_{\varphi,s},s) \notag\\
     & \stackrel{(a)}{\ge} B(\xi^1, s, \theta) V^2_t(\delta_0,s)
   + (1-B(\xi^1, s, \theta)) V^2_t(\xi^1|_{\theta,s},s) \notag \\
     & = H_t(\xi^1,s,\theta),
     \label{eq:H-ineq}
  \end{align}
where $(a)$ holds due to Properties~\ref{prop:6} and~\ref{prop:7} and~\eqref{eq:induction-2}. Then, we have from~\eqref{eq:dp-1-fin-e}
\begin{align}
V^1_t(\pi^1,s) &= \min_{\varphi \colon \ALPHABET X \to \ALPHABET U} 
          \Big\{ \Lambda (\pi^1, \varphi) +
       H_t(\pi^1, s, \varphi) 
      \Big\} \notag\\
      &\stackrel{(b)}{\ge} \min_{\theta \colon \ALPHABET X \to \ALPHABET U} 
          \Big\{ \Lambda (\xi^1, \theta) +
       H_t(\xi^1, s, \theta)
      \Big\} \notag\\
      &= V^1_t(\xi^1,s), \label{eq:induction-1} 
\end{align}
where $(b)$ holds due to Property~\ref{prop:6} and~\eqref{eq:H-ineq} and since inequality is preserved in point-wise minimization. This completes the induction step.

In order to show (V2), note that $\pi^1_t \succeq_a \pi^1_t$ trivially. Let
$\varphi_t$ be the optimal prescription at $\pi^1_t$. If $\varphi_t \in
\SQC(c)$, then we are done. If not, define $\theta_t= \mathcal T^{(\pi^1_t,
\pi^1_t)}_c \varphi_t$ where the transformation $\mathcal T^{(\pi^1_t,
\pi^1_t)}_c$ is introduced in Lemma~\ref{lemma:pres}. 
Then, the argument in~\eqref{eq:induction-1} (with $\xi^1$ replaced by
$\pi^1$) also implies (V2). Furthermore, (V3) follows from
Property~\ref{prop:1}.
\end{proof}

\subsection{Proof of Theorem~\ref{thm:SQC}}

\subsubsection*{Proof of Part~1}

Properties~\ref{prop:2} and~\ref{prop:3} imply that for
all~$t$, $\Pi^2_t$ is $\ASU(0)$. Therefore, by Property~\ref{prop:1}, the
optimal estimate $\hat E_t = 0$. Recall that $\hat E_t = \hat X_t - Z_t$.
Thus, $\hat X_t = Z_t$. This proves the first part of
Theorem~\ref{thm:SQC}.

\subsubsection*{Proof of Parts~2 and 3}

Lemmas~\ref{lemma:irrelevant_info_Tx} and~\ref{lemma:U_hat_E} imply that there
is no loss of optimality in restricting attention to transmitters of the
form
\begin{equation}\label{eq:transmit_E}
U_t = \tilde g_t(E_t, Y_{0:t-1}).
\end{equation}
Part~1 implies that there is no loss of optimality in restricting attention to
estimation strategies of the form~\eqref{eq:rx-a}. So, we assume that the
transmission and estimation strategies are of these forms.

Since the estimation strategy is fixed, Problem~\ref{prob:finite} reduces to a
single agent optimization problem. $\{(E_t, S_{t-1})\}_{t \ge 0}$ is an
information state for this single agent optimization problem for the following
reasons:
\begin{enumerate}
  \item Eq.~\eqref{eq:dynamics-2a} implies that $\{E_t\}_{t \ge 0}$ is a
    controlled Markov process controlled by $R_t$. Moreover, for any
    realization $(e_{0:t}, u_{0:t}, s_{0:t-1})$ of $(E_{0:t}, U_{0:t},
    S_{0:t-1})$, we have
    \begin{align*}
      \PR(R_t = 0 | e_{0:t}, u_{0:t}, s_{0:t-1})
      &= \sum_{s' \in \ALPHABET S} Q_{ss'} p(s', u)
      \\
      &= \PR(R_t = 0 | u_t, s_{t-1}).
    \end{align*}
    Combining the two we get that
    \[
      \PR(E_{t+1}, S_t | e_{0:t}, u_{0:t}, s_{0:t-1}) =
      \PR(E_{t+1}, S_t | e_{t}, u_{t}, s_{t-1}).
    \]
      
  \item Using~\eqref{eq:e-update}, the conditional expected per step cost may
    be written as
    \begin{align}\label{eq:per-step_cost}
      \hskip 2em & \hskip -2em
      \EXP[ \lambda(U_t) + d(E^+_t) | e_{0:t}, u_{0:t}, s_{0:t-1}]
      \notag \\
      &= \lambda(u_t) + \EXP [ p(S_t, u) d(e) | S_{t-1} = s ].
    \end{align}
\end{enumerate}
Thus, the optimization problem at the transmitter is an MDP with information
state $(E_t, S_{t-1})$. Therefore, from Markov decision theory, there is no
loss of optimality in restriction attention to Markov strategies of the form
$f_t(E_t, S_{t-1})$. The optimal strategies of this form are given by the
dynamic program of Part~3.

\section{Proof of Theorem~\ref{thm:SQC-infin}}\label{app:AR-infin-proof}
\subsection{Proof of Part~1)}
The structure of Theorem~~\ref{thm:SQC} holds the the infinite horizon setup as well. The main idea is to use forward induction to show that the optimal estimate for the error process, $\hat E_t = 0$. The structure of optimal $\hat X_t$ is then derived by using the change of variable introduced in Section~\ref{sec:error_process}.

\subsection{Some preliminary properties}
We prove the following properties, which will be used to establish the
existence of the solution to the dynamic program~~\eqref{eq:dp-J1-infin}. Note
that the per-step cost $c(e,s,u)$ given in~\eqref{eq:per-step_cost} can be
rewritten as $c(e,s,u) \DEFINED (1-\beta) (\lambda(u) + \sum_{s_t \in
\ALPHABET S} Q_{s s_t}p(s_t,u)d(e))$. 

Our model satisfies the following properties (\cite[Assumtions~4.2.1,4.2.2]{LermaLasserre}).
\begin{proposition}\label{prop:p1-p3}
  Under Assumption~\ref{assump:bounded-dist}, for any $\lambda(\cdot) \ge 0$,
  the following conditions of~\cite{LermaLasserre} are satisfied:
  \begin{enumerate}
    \item[(P1)] The per-step cost $c(e,s,u) $ is lower semi-continuous\footnote{A function is lower semi-continuous if its lower level sets are closed.}, bounded from below and inf-compact\footnote{A function $v: \ALPHABET X \times \ALPHABET U \to \reals$ is said to be \textit{inf-compact} on $\ALPHABET X \times \ALPHABET U$ if, for every $x \in \ALPHABET X$ and $r \in \reals$, the set $\{u \in \ALPHABET U\,:\, v(x,u) \le r\}$ is compact.} on $\reals \times \{0,1\}$, i.e., for all $e, r \in \reals$, the set $\{u \in \ALPHABET U\,:\, c(e,s,u) \le r\}$ is compact.
    \item[(P2)] For every $u \in \{0,1\}$, the transition kernel from $(E_t,S_{{t-1}})$ to $(E_{t+1}, S_t)$ is strongly continuous\footnote{A controlled transition probability kernel $P(\cdot\,|\, x,u): \ALPHABET X \rightarrow [0,1]$ is said to be \textit{strongly continuous} if for any bounded measurable function $v$ on $\ALPHABET X$ the function $v'$ on $\ALPHABET X \times \ALPHABET U$, $v'(x,u) \DEFINED \EXP^P [v\,|\, x,u]$, $x \in \ALPHABET X$, $u \in \ALPHABET U$, is continuous and bounded.}.
    \item[(P3)] There exists a strategy for which the value function is finite. 
  \end{enumerate}
\end{proposition}
\begin{proof}
(P1) is true because of the following reasons. The action set $\ALPHABET U = \ALPHABET U$ is either finite or uncountable and the per-step cost $c(\cdot,s,u)$ is continuous on $\reals$ (and hence lower semi-continuous), and is non-negative. Finally, when $\ALPHABET U$ is finite, all subsets of $\ALPHABET U$ are compact.
When $\ALPHABET U$ is uncountable, all closed subsets of $\ALPHABET U$ are compact.

To check (P2), note the following fact (\cite[Example~C.6]{LermaLasserre}).

\begin{description}\label{fact:strong-cont-density}
\item[Fact~1] Let $P$ be a stochastic kernel and suppose that there is a $\sigma$-finite measure $\phi$ on $\ALPHABET X$ such that, for every $y \in \ALPHABET X$, $\PR(\cdot|y)$ has a density $p(\cdot|y)$ with respect to $\phi$, that is,
\[
  P(B|y) = \int_{B} p(x|y)\phi(dx), \quad \forall B \in \mathcal B(\ALPHABET X), y \in \ALPHABET X.
\]
If $p(x|\cdot)$ is continuous on $\ALPHABET X$ for every $x \in \ALPHABET X$, then $P$ is strongly continuous.
\end{description}

(P2) is true for the following reasons. Let $P$ denote the transition kernel
from $(E_t,S_{{t-1}})$ to $(E_{t+1}, S_t)$. Then for any Borel subset $B$ of
$\mathcal B(\ALPHABET X)$, 
\begin{multline*}
  P(E_{t+1} \in B \,|\, E_t = e, S_{t-1} = s, U_t = u) \\
  = (1-p(s,u))\int_{B} \mu(w) dw + p(s,u) \int_{B} \mu(w-ae) dw.
\end{multline*}
Then, according to Fact~1, $P$ is strongly continuous since the real line $\reals$ with Lebesgue measure $\phi(dx) = dx$ is $\sigma$-finite and since the density $\mu$ is continuous on $\reals$. 

(P3) is true due to Assumption~\ref{assump:bounded-dist}.
\end{proof}

\subsection{Proofs of Parts~2) and~3)}
For ease of exposition, we assume $\ALPHABET X = \reals$. Similar argument
works for $\ALPHABET X = \integers$. We fix the optimal estimator with the
Kalman-filter like structure~\eqref{eq:rx-a} and identify the best performing
transmitter, which is a centralized optimization problem. For the discounted
setup, one expects that the optimal solution is given by the fixed point of
the dynamic program~\eqref{eq:dp-J1-infin} (similar
to~\eqref{eq:dp-J0}--\eqref{eq:dp-J1}). However, it is not obvious that there
exists a fixed point of~\eqref{eq:dp-J1-infin} because the distortion
$d(\cdot)$ is unbounded.

Define the operator $\mathcal B$ given as follows: for any given $s \in \ALPHABET S$ and function $v: \reals \times \ALPHABET S \rightarrow \reals$
\begin{align*}
[\mathcal B v](e,s) &\DEFINED \min_{u \in \ALPHABET U} \EXP[c(e,s,u) + v(E_{t+1},S_t)\\
&\hskip 8em \,|\, E_t = e, S_{t-1} = s].
\end{align*}
Then~\eqref{eq:dp-J1-infin} can be expressed in terms of the operator $\mathcal B$ as follows:
\begin{align*}
  [\mathcal B J_\beta](e,s) &\DEFINED \min_{u \in \ALPHABET U} \EXP[c(e,s,u) + J_\beta(E_{t+1},S_t)\\
  &\hskip 8em \,|\, E_t = e, S_{t-1} = s].
\end{align*}
Then, the proof of Parts~2) and~3) of the theorem follows directly from~\cite{LermaLasserre}. In particular, we have the following proposition, where the first part follows from~\cite[Theorem~4.2.3]{LermaLasserre} and the second part follows from~\cite[Lemma~4.2.8]{LermaLasserre}.
\begin{proposition}\label{prop:DC_infin_DP}
  Under (P1)--(P3), there exist fixed point solutions $J_\beta$
  of~\eqref{eq:dp-J1-infin}. Let $J^*_\beta: \ALPHABET X \rightarrow \reals$ denote
  the smallest such fixed point and $\tilde f^*_\beta(e,s)$ denote the argmin
  of the right hand side of~\eqref{eq:dp-J1-infin} for $J_\beta = J^*_\beta$.
  Then,
  \begin{enumerate}
    \item $\tilde f^*_\beta$ is the optimal strategy for Problem~\ref{prob:infinite} with $\beta \in (0,1)$. 
    \item Let $J^{(0)}_\beta = 0$ and define $J^{(n+1)}_\beta \DEFINED \mathcal B J^{(n)}_\beta$. 
      Then, $J^*_\beta = \lim_{n \rightarrow \infty} J^{(n)}_\beta$.
  \end{enumerate}
\end{proposition}

\subsection{Properties of the value function}
We can apply the \emph{vanishing discount approach} and show that the result for the long-term average cost is obtained as a limit of those in the discounted setup, as $\beta \uparrow 1$. 

Our model satisfies the following conditions~~\cite{Sennott:book,LermaLasserre}.
\begin{proposition}
  \label{prop:SEN}
  Under Assumption~\ref{assump:bounded-dist}, for any $\lambda(\cdot) \ge 0$, the value function $J_\beta$, as given by~\eqref{eq:dp-J1-infin}, satisfies the following conditions of~\cite{Sennott:book,LermaLasserre}: for any $s \in \ALPHABET S$, and $\lambda \in \reals_{\ge 0}$, 
   \begin{enumerate}
    \item[(S1)] there exists a reference state $e_0 \in \ALPHABET X$ and a
      non-negative scalar $M_s$ such that $J^*_\beta(e_0, s) <
      M_s$ for all $\beta \in (0,1)$.
    \item[(S2)] Define $h_\beta(e,s) = (1-\beta)^{-1}[
        J_\beta(e,s) - J_\beta(e_0,s)]$. There exists a function
        $K_s : \ALPHABET X  \to \reals$ such that $h_\beta(e,s) \le
        K_s(e)$ for all $e \in \ALPHABET X$ and $\beta \in (0,1)$. 
    \item[(S3)] There exists a non-negative (finite) constant $L_\lambda$
      such that $-L_s \le h_\beta(e,s)$ for all $e \in \ALPHABET X$
      and $\beta \in (0,1)$. 
  \end{enumerate}
  Therefore, if $\tilde f^*_\beta$ denotes an optimal strategy for $\beta \in (0,1)$,
  and $\tilde f^*_1$ is any limit point of $\{\tilde f^*_\beta\}$, then $\tilde f^*_1$ is optimal for
  $\beta =1$.
\end{proposition}

\begin{proof}
  We prove the proposition for $\ALPHABET X = \reals$. Similar argument holds
  for $\ALPHABET X = \integers$.

  Let $\tilde J^{(0)}_\beta(e,s)$ denote the value function of the `always
  transmit with maximum power' strategy. According to
  Assumption~\ref{assump:bounded-dist}, $M_s = \tilde J^{(0)}_\beta(0,s) <
  \infty$. Hence, (S1) is satisfied with $e_0 = 0$ and $M_s = \tilde
  J^{(0)}_\beta(0,s)$.

  Since not transmitting is optimal at state 0 (because the transmission
  strategy is $\SQC$ about 0), we have
  \begin{align*}
    J^*_\beta(0,s) &= \beta \sum_{s_t \in \ALPHABET S} Q_{ss_t} \int_{\reals} \mu(w) J^*_\beta(w,s_t)dw .
  \end{align*}
  Let $\tilde J^{(1)}_\beta(e,s)$ denote the value function of the strategy
  that transmits with power level $u = \varphi(e)$ at time 0 and follows the
  optimal strategy from then on. Then
  \begin{align}
    &\tilde J^{(1)}_\beta(e,s) = (1-\beta) [\lambda(\varphi(e)) + \sum_{s_t \in \ALPHABET S} Q_{s s_t} p(s_t,\varphi(e))d(e)] \notag\\
    &\hskip 6em + \beta \sum_{s_t \in \ALPHABET S} Q_{s s_t} \int_{\reals} \mu(w) J^*_\beta(w,s_t)dw \notag\\
    &= (1-\beta)[\lambda(\varphi(e)) + \sum_{s_t \in \ALPHABET S} Q_{s s_t}p(s_t,\varphi(e))d(e)] \notag\\ %+ J^*_\beta(0,s)
    & \hskip 16em + J^*_\beta(0,s).
    \label{eq:proofSEN}
  \end{align}
  Since $J^*_\beta(e,s) \le \tilde J^{(1)}_\beta(e,s)$ and $J^*_\beta(0,s) \ge
  0$, from~\eqref{eq:proofSEN} we get that $(1-\beta)^{-1} [J^*_\beta(e,s) -
  J^*_\beta(0,s)] \le \lambda(\varphi(e)) + \sum_{s_t \in \ALPHABET S} Q_{s
  s_t} p(s_t,\varphi(e))d(e)$. Hence (S2) is satisfied with $K_s(e) =
  \lambda(\varphi(e)) + \sum_{s_t \in \ALPHABET S} Q_{s
  s_t}p(s_t,\varphi(e))d(e)$.

  According to~\cite[Theorem~1]{JC-AM-mono-TAC2019}, the value function
  $J^*_\beta$ is even and quasi-convex and hence $J^*_\beta(e,s) \ge
  J^*_\beta(0,s)$. Hence (S3) is satisfied with $L_s = 0$.
\end{proof}

\subsection{Proof of Part~4)}
The result for Part~4) of the theorem for $\ALPHABET X = \integers$ follows from~\cite[Theorem~7.2.3]{Sennott:book} and for $\ALPHABET X = \reals$ the result of Part~4) follows from~\cite[Theorem~5.4.3]{LermaLasserre}.

\section{Proof of Theorem~\ref{thm:DNC}}\label{app:proof-thm_DNC}
For ease of notation, we use $\tau$ instead of $\tau^{(1)}$.
Let $\mathcal F_0$ denote the event $\{E^+_{-1} = 0, S_{-1} = s^\circ\}$ and 
$\mathcal F_\tau$ denote the event $\{E^+_{\tau-1} = 0, S_{\tau-1} = s^\circ
\}$.

First note that from~\eqref{eq:M-tau} we can write
\begin{equation}\label{eq:beta-tau}
  \EXP[\beta^{\tau} \,|\, \mathcal F_0] = 1-(1-\beta)M^{(\mathbf k)}_\beta.
\end{equation}

Now consider
\begin{align}
  C^{(\mathbf k)}_\beta &= (1-\beta) \EXP \Big[\sum_{t=0}^\infty
    \beta^t (\lambda(U_t) + d(E^+_t))\Bigm| \mathcal F_0 \Big]
    \notag \\
  &= (1-\beta)\EXP \Big[ \sum_{t=0}^{\tau-1} \beta^t (\lambda(U_t) +
  d(E^+_t))\Bigm| \mathcal F_0 \Big] \notag\\
  & \quad + (1-\beta) \EXP \Big[\beta^\tau \sum_{t=\tau}^\infty \beta^{t-\tau}
  (\lambda(U_t) + d(E^+_t)) \Bigm| \mathcal F_0 \Big] \notag\\
  &\stackrel{(a)}{=} (1-\beta) L^{(\mathbf k)}_\beta  \notag 
  + (1-\beta) \EXP[\beta^\tau\,|\, \mathcal F_0]  \notag \\
  & \quad \times 
  \EXP \Big[ \sum_{t=\tau}^\infty \beta^{t-\tau} (\lambda(U_t) + d(E^+_t)) 
  \Bigm| \mathcal F_\tau \Big] \notag\\
  & \stackrel{(b)}{=} (1-\beta) L^{(\mathbf k)}_\beta + [ 1 - (1 - \beta) M^{(\mathbf k)}_\beta ] C^{(\mathbf k)}_\beta, \label{eq:Ck_proof}
\end{align}
where the first term of $(a)$ uses the definition of $L^{(\mathbf k)}_\beta$
as given by~\eqref{eq:L-tau} and the second term of $(a)$ uses strong Markov
property; $(b)$ uses~\eqref{eq:beta-tau} and time-homogenity. Rearranging terms in~\eqref{eq:Ck_proof} we get~\eqref{eq:Ck}.

\bibliographystyle{IEEEtran}
\bibliography{IEEEabrv,markovTAC_ref} 

\begin{thebibliography}{10}
\providecommand{\url}[1]{#1}
\csname url@samestyle\endcsname
\providecommand{\newblock}{\relax}
\providecommand{\bibinfo}[2]{#2}
\providecommand{\BIBentrySTDinterwordspacing}{\spaceskip=0pt\relax}
\providecommand{\BIBentryALTinterwordstretchfactor}{4}
\providecommand{\BIBentryALTinterwordspacing}{\spaceskip=\fontdimen2\font plus
\BIBentryALTinterwordstretchfactor\fontdimen3\font minus
  \fontdimen4\font\relax}
\providecommand{\BIBforeignlanguage}[2]{{%
\expandafter\ifx\csname l@#1\endcsname\relax
\typeout{** WARNING: IEEEtran.bst: No hyphenation pattern has been}%
\typeout{** loaded for the language `#1'. Using the pattern for}%
\typeout{** the default language instead.}%
\else
\language=\csname l@#1\endcsname
\fi
#2}}
\providecommand{\BIBdecl}{\relax}
\BIBdecl

\bibitem{XuHes2004a}
Y.~Xu and J.~P. Hespanha, ``Optimal communication logics in networked control
  systems,'' in \emph{Proc. IEEE Conference on Decision and Control}, 2004, pp.
  3527--3532.

\bibitem{ImerBasar}
O.~C. Imer and T.~Ba{\c{s}}ar, ``Optimal estimation with limited
  measurements,'' in \emph{Proc. IEEE Conference on Decision and Control and
  European Control Conference}, 2005, pp. 1029--1034.

\bibitem{Rabi2012}
M.~Rabi, G.~Moustakides, and J.~Baras, ``Adaptive sampling for linear state
  estimation,'' \emph{SIAM Journal on Control and Optimization}, vol.~50,
  no.~2, pp. 672--702, 2012.

\bibitem{LipsaMartins:2011}
G.~M. Lipsa and N.~C. Martins, ``Remote state estimation with communication
  costs for first-order {LTI} systems,'' \emph{{IEEE} Trans. Autom. Control},
  vol.~56, no.~9, pp. 2013--2025, 2011.

\bibitem{NayyarBasarTeneketzisVeeravalli:2013}
A.~Nayyar, T.~Ba{\c{s}}ar, D.~Teneketzis, and V.~V. Veeravalli, ``Optimal
  strategies for communication and remote estimation with an energy harvesting
  sensor,'' \emph{{IEEE} Trans. Autom. Control}, vol.~58, no.~9, pp.
  2246--2260, 2013.

\bibitem{MH2017}
A.~Molin and S.~Hirche, ``Event-triggered state estimation: An iterative
  algorithm and optimality properties,'' \emph{{IEEE} Trans. Autom. Control},
  vol.~62, no.~11, pp. 5939--5946, Nov 2017.

\bibitem{JC_AM_IFAC16}
J.~Chakravorty and A.~Mahajan, ``Remote state estimation with packet drop,'' in
  \emph{Proc. of 6th {IFAC} Workshop on Distributed Estimation and Control in
  Networked Systems}, Sep 2016, pp. 7--12.

\bibitem{LipsaMArtins2009}
G.~M. Lipsa and N.~C. Martins, ``Optimal state estimation in the presence of
  communication costs and packet drops,'' in \emph{2009 47th Annual Allerton
  Conference on Communication, Control, and Computing (Allerton)}, Sept 2009,
  pp. 160--169.

\bibitem{JC-JS-AM-ACC17}
J.~Chakravorty, J.~Subramanian, and A.~Mahajan, ``Stochastic approximation
  based methods for computing the optimal thresholds in remote-state estimation
  with packet drops,'' in \emph{Proc. of IEEE American Control Conference}, May
  2017, pp. 462--467.

\bibitem{gaoEtal18}
X.~Gao, E.~Akyol, and T.~Ba\c{s}ar, ``Optimal communication scheduling and
  remote estimation over an additive noise channel,'' \emph{Automatica},
  vol.~88, pp. 57--69, 2018.

\bibitem{JC_AM_ISIT17}
J.~Chakravorty and A.~Mahajan, ``Structure of optimal strategies for remote
  estimation over {G}ilbert-{E}lliott channel with feedback,'' in \emph{Proc.
  of the IEEE International Symposium on Information Theory}, Jun 2017, pp.
  1272--1276.

\bibitem{Renetal2017}
X.~Ren, J.~Wu, K.~H. Johansson, G.~Shi, and L.~Shi, ``Infinite horizon optimal
  transmission power control for remote state estimation over fading
  channels,'' \emph{{IEEE} Trans. Autom. Control}, vol.~63, no.~1, pp. 85--100,
  Jan 2018.

\bibitem{MH2012}
A.~Molin and S.~Hirche, ``An iterative algorithm for optimal event-triggered
  estimation,'' in \emph{4th IFAC Conference on Analysis and Design of Hybrid
  Systems (ADHS'12)}, 2012, pp. 64--69.

\bibitem{Gatsisetal2014}
K.~Gatsis, A.~Ribeiro, and G.~J. Pappas, ``Optimal power management in wireless
  control systems,'' \emph{{IEEE} Trans. Autom. Control}, vol.~59, no.~6, pp.
  1495--1510, June 2014.

\bibitem{JC_AM_TAC17}
J.~Chakravorty and A.~Mahajan, ``Fundamental limits of remote estimation of
  autoregressive {M}arkov processes under communication constraints,''
  \emph{{IEEE} Trans. Autom. Control}, vol.~62, no.~3, pp. 1109--1124, March
  2017.

\bibitem{Heetal}
L.~He, J.~Chen, and Y.~Qi, ``Event-based state estimation: Optimal algorithm
  with generalized closed skew normal distribution,'' \emph{to appear in IEEE
  Transactions on Automatic Control}, pp. 1--8, 2018.

\bibitem{Chenetal2017}
W.~Chen, J.~Wang, D.~Shi, and L.~Shi, ``Event-based state estimation of hidden
  {M}arkov models through a {G}ilbert-{E}lliott channel,'' \emph{{IEEE} Trans.
  Autom. Control}, vol.~62, no.~7, pp. 3626--3633, July 2017.

\bibitem{GoldsmithVaraiya1996}
A.~J. Goldsmith and P.~P. Varaiya, ``Capacity, mutual information, and coding
  for finite-state {M}arkov channels,'' \emph{{IEEE} Trans. Inf. Theory},
  vol.~42, no.~3, pp. 868--886, May 1996.

\bibitem{YangEtAl2005}
S.~Yang, A.~Kavcic, and S.~Tatikonda, ``Feedback capacity of finite-state
  machine channels,'' \emph{{IEEE} Trans. Inf. Theory}, vol.~51, no.~3, pp.
  799--810, March 2005.

\bibitem{WalrandVaraiya:1983}
J.~C. Walrand and P.~Varaiya, ``Optimal causal coding-decoding problems,''
  \emph{{IEEE} Trans. Inf. Theory}, vol.~29, no.~6, pp. 814--820, Nov. 1983.

\bibitem{NMT:partial-history-sharing}
A.~Nayyar, A.~Mahajan, and D.~Teneketzis, ``Decentralized stochastic control
  with partial history sharing: A common information approach,'' \emph{{IEEE}
  Trans. Autom. Control}, vol.~58, no.~7, pp. 1644--1658, jul 2013.

\bibitem{JC-AM-mono-TAC2019}
J.~Chakravorty and A.~Mahajan, ``Sufficient conditions for the value function
  and optimal strategy to be even and quasi-convex,'' \emph{to appear in IEEE
  Transactions on Automatic Control}, 2019.

\bibitem{Huang_Dey_intermittentKF}
M.~Huang and S.~Dey, ``Stability of {K}alman filtering with {M}arkovian packet
  losses,'' \emph{Automatica}, vol.~43, no.~4, pp. 598--607, 2007.

\bibitem{Rohretal_intermittentKF}
E.~R. Rohr, D.~Marelli, and M.~Fu, ``{K}alman filtering with intermittent
  observations: On the boundedness of the expected error covariance,''
  \emph{{IEEE} Trans. Autom. Control}, vol.~59, no.~10, pp. 2724--2738, Oct
  2014.

\bibitem{wuetal2018}
J.~Wu, G.~Shi, B.~D.~O. Anderson, and K.~H. Johansson, ``Kalman filtering over
  {G}ilbert-{E}lliott channels: Stability conditions and critical curve,''
  \emph{{IEEE} Trans. Autom. Control}, vol.~63, no.~4, pp. 1003--1017, April
  2018.

\bibitem{LermaLasserre}
O.~H. Lerma and J.~B. Lasserre, \emph{Discrete-time {M}arkov control processes
  : basic optimality criteria}.\hskip 1em plus 0.5em minus 0.4em\relax
  Springer, 1996.

\bibitem{JS-AM-RMC-arxiv}
J.~{Subramanian} and A.~{Mahajan}, ``{Renewal Monte Carlo: Renewal theory based
  reinforcement learning},'' \emph{arXiv: 1804.01116}, Apr. 2018.

\bibitem{spall1992multivariate}
J.~C. Spall, ``Multivariate stochastic approximation using a simultaneous
  perturbation gradient approximation,'' \emph{{IEEE} Trans. Autom. Control},
  vol.~37, no.~3, pp. 332--341, Mar 1992.

\bibitem{MaryakChin}
J.~L. Maryak and D.~C. Chin, ``Global random optimization by simultaneous
  perturbation stochastic approximation,'' \emph{{IEEE} Trans. Autom. Control},
  vol.~53, no.~3, pp. 780--783, 2008.

\bibitem{KK1972}
V.~Katkovnik and Y.~Kulchitsky, ``Convergence of a class of random search
  algorithms,'' \emph{Automation and Remote Control}, vol.~33, no.~8, pp.
  1321--1326, 1972.

\bibitem{Bhatnagar2013}
S.~Bhatnagar, H.~Prasad, and L.~Prashanth, \emph{Stochastic Approximation
  Algorithms}.\hskip 1em plus 0.5em minus 0.4em\relax London: Springer, 2013,
  pp. 17--28.

\bibitem{gilbert1960}
E.~N. Gilbert, ``Capacity of a burst-noise channel,'' \emph{Bell System
  Technical Journal}, vol.~39, no.~5, pp. 1253--1265, 1960.

\bibitem{elliott1963}
E.~O. Elliott, ``Estimates of error rates for codes on burst-noise channels,''
  \emph{Bell System Technical Journal}, vol.~42, no.~5, pp. 1977--1997, 1963.

\bibitem{kingma_ba15}
D.~P. Kingma and J.~Ba, ``Adam: {A} method for stochastic optimization,''
  \emph{arxiv: 1412.6980}, Jan 2017.

\bibitem{Tameretalacc16}
X.~Gao, E.~Akyol, and T.~Ba{\c{s}}ar, ``On remote estimation with multiple
  communication channels,'' in \emph{Proc. American Control Conference}, July
  2016, pp. 5425--5430.

\bibitem{KumarVaraiya:1986}
P.~R. Kumar and P.~Varaiya, \emph{Stochastic Systems: Estimation,
  Identification and Adaptive Control}.\hskip 1em plus 0.5em minus 0.4em\relax
  NJ, USA: Prentice-Hall, Inc., 1986.

\bibitem{Sennott:book}
L.~I. Sennott, \emph{Stochastic dynamic programming and the control of queueing
  systems}.\hskip 1em plus 0.5em minus 0.4em\relax New York, NY, USA: Wiley,
  1999.

\bibitem{Puterman:1994}
M.~Puterman, \emph{{M}arkov decision processes: Discrete Stochastic Dynamic
  Programming}.\hskip 1em plus 0.5em minus 0.4em\relax John Wiley and Sons,
  1994.

\end{thebibliography}

\newpage

\section{Proof of Lemma~\ref{lemma:F1-F2}}\label{app:proof-F1-F2}
Given any realization $(x_{0:T}, s_{0:T}, y_{0:T}, u_{0:T})$ of the system
variables $(X_{0:T}, S_{0:T}, Y_{0:T}, U_{0:T})$, consider
\begin{align}
  \pi^1_{t+1}(x_{t+1}) &= \PR(x_{t+1}|s_{0:t}, y_{0:t}) \notag \\
  &= \sum_{x_t \in \ALPHABET X} \PR(x_{t+1} | x_t) \PR(x_t | s_{0:t}, y_{0:t})
  \notag \\
  &= \sum_{x_t \in \ALPHABET X} P_{x_t x_{t+1}} \pi^2_t(x_t) = \pi^2_t P
  \label{eq:F1-update}
\end{align}
which is the expression for $F^1(\cdot)$.

For $F^2$, we consider the following two cases separately. First, given any realization $(x_{0:T}, s_{0:T}, y_{0:T}, u_{0:T})$ of the system
variables $(X_{0:T}, S_{0:T}, Y_{0:T}, U_{0:T})$, consider
\begin{equation}
  \pi^2_t(x) = \PR(X_t = x | s_{0:t}, y_{0:t}) %= \IND_{\{ x = y_t \}}.
  \label{eq:pi-2-1}
\end{equation}
\subsubsection{Case~1: $y_t \in \ALPHABET X$}
For $y_t \in \ALPHABET X$, we have
\begin{equation}
  \label{eq:pi-2-2}
\end{equation}
\subsubsection{Case~2: $y_t = \BLANK$}
In this case we have
\begin{align}
  \pi^2_t(x) &= \PR(X_t = x | s_{0:t}, y_{0:t})  \notag \\
  &= \frac { \PR(X_t = x, Y_t = y_t | s_{0:t}, y_{0:t-1}) }
           { \PR(Y_t = y_t | s_{0:t}, y_{0:t-1}) }.
  \label{eq:pi-2-3}
\end{align}
Now, consider the numerator of~\eqref{eq:pi-2-3}.
\begin{align}
  &\PR(X_t = x_t,Y_t = \BLANK\,|\, s_{0:t}, y_{0:t-1}) \notag\\
  & \hskip 2em = \PR(Y_t = \BLANK\,|\, s_{0:t},y_{0:t-1},x_t)
    \PR(x_t\,|\,s_{0:t},y_{0:t-1}) \notag\\
  & \hskip 2em \stackrel{(a)}{=} \sum_{u \in \ALPHABET U}
    \IND_{\{\varphi_t(x_t) = u\}} p(s_t, \varphi_t(x_t)))\pi^1_t(x_t),
  \label{eq:num}
\end{align}
where in $(a)$ we use~\eqref{eq:channel} and the fact that $\PR(x_t\,|\,s_{0:t},y_{0:t-1}) = \PR(x_t\,|\,s_{0:t-1},y_{0:t-1})$ because the channel state is independent of the source.

The denominator of~\eqref{eq:pi-2-3} is obtained by averaging the numerator
over $x_t \in \ALPHABET X$.

\section{Proof of Theorem~\ref{thm:monotonic-threshold-fin}}
\label{subsec:monotonic-threshold-fin}
We first note that Assumption~\ref{assump:sto-mono} implies the following.
\begin{lemma}\label{lemma:expectation_stoc_monotone}
  \cite{Puterman:1994}
  For any weakly increasing (respectively weakly decreasing) function
  $h^\circ$ we have
  \[
    \EXP[h^\circ(S_t)\,|\, S_{t-1} = s]
  \]
  is weakly increasing (respectively weakly decreasing) in $s$.
\end{lemma}

\begin{proof}[Theorem~\ref{thm:monotonic-threshold-fin}, Parts~1) and~3)]
  The results follow directly from~\cite[Theorem 1]{JC-AM-mono-TAC2019}. We
  verify conditions (C1)--(C5) of~\cite[Theorem 1]{JC-AM-mono-TAC2019} in
  Section~\ref{app:proof_SQC}. 
\end{proof}

\begin{remark}\label{rem:j_beta_EQ}
  Using similar argument as in finite horizon case, one can show that the for
  discounted case infinite horizon, with discount factor $\beta \in (0,1)$ and
  any $s \in \ALPHABET S$, the optimal value function $J^*_\beta(e,s)$ and the
  optimal rules $\tilde f^*_\beta(e,s)$ are even and quasi-convex in $e$.
\end{remark}
 
\begin{proof}[Theorem~\ref{thm:monotonic-threshold-fin}, Part~2)]
  First we note that we can rewrite the right hand side of~\eqref{eq:dp-J1} as
  follows:
  \[
    \bar H_t(e,s,u) \DEFINED \lambda(u) + \sum_{s_t \in \ALPHABET S} Q_{s s_t} [ p(s_t,u)d(e) + \bar J_t(e,s_t,u)],
  \]
  where $\bar J_t(e,s,u) \DEFINED \EXP_W [J_{t+1}(aE_t+W_t,s)\,|\, E_t = e, U_t = u]$. 

  We prove the result of Part~2) by backward induction. The statement is
  trivially true for $t = T+1$ as $J_{T+1} = 0$. Assume that for all $e \in
  \ALPHABET X$, $J_{t+1}(e,s)$ is decreasing in $s$. Since $\bar J_t(e,s,u)$
  is a weighted average of $J_{t+1}(E_{t+1},s)$ with non-negative weights, we
  have that for all $e \in \ALPHABET X$, $u \in \ALPHABET U$, $\bar
  J_t(e,s,u)$ is weakly decreasing in $s$. By assumption, $p(s,u)$ is also
  decreasing in $s$. Thus, for any $(e,u)$, $p(s,u)d(e)+\bar J_t(e,s)$ is
  decreasing in $s$. Thus by Lemma~\ref{lemma:expectation_stoc_monotone}, for
  any $(e,s)$, $\bar H_t(e,s,u)$ is decreasing in $s$. Since the pointwise
  minimum of decreasing functions is decreasing, from~\eqref{eq:dp-J1} we have
  that $J_t(e,s)$ is decreasing in $s$. This completes the induction step.
\end{proof}

\subsection{Verification of conditions (C1)--(C5) of \cite[Theorem 2]{JC-AM-mono-TAC2019}}
\label{app:proof_SQC}
We prove the result for $\ALPHABET X = \reals$. Similar argument holds for
$\ALPHABET X = \integers$. To prove the result, we show that dynamic
program~\eqref{eq:dp-J0}--\eqref{eq:dp-J1} satisfies conditions (C1)--(C5)
of~\cite[Theorem 1]{JC-AM-mono-TAC2019}. We use the notation $\SQC(0)$ for
even and quasi-convex. To prove the lemma, we first define the following:
\begin{itemize}
  \item The \emph{per-step cost function} $c$ as follows: for any $e \in
    \reals; s \in \ALPHABET S$ and $u \in \ALPHABET U$,
    \begin{equation}\label{eq:c_fin}
      c(e,s,u) \DEFINED \lambda(u) + \sum_{s_t \in \ALPHABET S} Q_{s s_t}p(s,u)d(e).
    \end{equation}

  \item Let $\tilde p(e^+\,|\, e,s;u)$ denote the \emph{conditional density}
    of $E_t = e^+$ given $E_t=e$, $S_{t-1}=s$ and $U_t = u$.
  \item Let $A_y \DEFINED \{e^+ \in \reals\,:\, e^+ < y\}$. 

  \item Define the function $M$ as follows: $M(y\,|\, e,s;u) \DEFINED 1-
    \int_{A_y} (\tilde p(e^+|e,s;u)+\tilde p(-e^+|e,s;u))de^+$.
\end{itemize}

For ease of reference, we restate the properties of the model:
\begin{enumerate}
  \item[(M0)] $p(s,0) = 1$ and $p(s,u_{\max}) \ge 0$.
  \item[(M1)] $\lambda(\cdot)$ is increasing with $\lambda(0) = 0$.
  \item[(M2)] $p(\cdot,\cdot)$ is decreasing in $s$ and $u$.
  \item[(M3)] $d(\cdot)$ is even and quasi-convex with $d(0) = 0$.
\end{enumerate}
In addition, we impose the following assumptions on the probability
density/mass function of the i.i.d.\@ process $\{W_t\}_{t \ge 1}$:
\begin{enumerate}
  \item[(M4)] The density $\mu$ of $W_t$ is even.
  \item[(M5)] $\mu(\cdot)$ is unimodal (i.e., quasi-concave).
\end{enumerate}

The following result follows from~\cite[Lemma~2]{JC-AM-mono-TAC2019}.
\begin{lemma} \label{lem:basic}
  Under \textup{(M4)} and \textup{(M5)}, for any $e,y \in \ALPHABET X_{\ge
  0}$, we have that
  \[
    \mu(y - e) \ge \mu (y + e)
  \]
\end{lemma}

An immediate implication of Lemma~\ref{lem:basic} are the following.
\begin{lemma} \label{lem:mono}
  Under \textup{(M4)} and \textup{(M5)}, for any $a \in \ALPHABET X$ and $e,y
  \in \ALPHABET X_{\ge 0}$, we have that
  \[
    a \big[ \mu(y - ae) - \mu (y + ae) \big] \ge 0.
  \]
\end{lemma}
\begin{proof}
  For $a \in \ALPHABET X_{\ge 0}$, from Lemma~\ref{lem:basic} we get that
  $\mu(y - ae) \ge \mu(y + ae)$. 
  For $a \in \ALPHABET X_{< 0}$, from Lemma~\ref{lem:basic} we get that
  $\mu(y + ae) \ge \mu(y - ae)$.
\end{proof}

We now state and verify conditions (C1)--(C5)
from~\cite[Theorem~2]{JC-AM-mono-TAC2019}.
\subsubsection{Condition~(C1)}
\emph{For any $s \in \ALPHABET S$ and $u \in \ALPHABET U$, the per-step cost
$c(e,s,u)$ is even and quasi-convex in $e$}.

This follows from~\eqref{eq:c_fin} and (M3).

\subsubsection{Condition~(C2)}
\emph{For any $s \in \ALPHABET S$ and $u \in \ALPHABET U$, $p(\cdot\,|\, \cdot,s;u)$ is even, i.e., for any $e^+, e \in \reals$}, 
\begin{equation}\label{eq:p_even}
  \tilde p(e^+\,|\, e,s;u) = \tilde p(-e^+\,|\, -e,s;u).
\end{equation}

This follows from (M4).

\subsubsection{Condition~(C3)}
\emph{For any $s \in \ALPHABET S$, $u \in \ALPHABET U$ and $y \in \reals_{\ge 0}$, $M(y\,|\, e,s;u)$ is (weakly) increasing in $e$ for all $e \in \reals_{\ge 0}$}. 

In order to do so, we arbitrarily fix $s \in \ALPHABET S$.

Consider any $y \in \reals_{\ge 0}$. Then,
\begin{align}
  &M(y\,|\, e,s;u) \notag\\
  &\hskip 2em = 1- \sum_{s_t \in \ALPHABET S} Q_{s s_t} p(s_t,u) \notag\\
  &\hskip 2em \times \Big[\int_{-\infty}^{y} (\mu(e^+-ae) + \mu(e^+ + ae)) de^+ \Big] \notag \\
  & \hskip 2em - \sum_{s_t \in \ALPHABET S} Q_{s s_t} (1-p(s_t,u)) \Big[\int_{-\infty}^y (\mu(e^+) + \mu(-e^+)) de^+\Big] \notag\\
  & \hskip 2em \stackrel{(a)}{=} 1 - \sum_{s_t \in \ALPHABET S} Q_{s s_t} p(s_t,u) [\mathcal M(y-ae) + \mathcal M(y+ae)] \notag \\
  &\hskip 2em - \sum_{s_t \in \ALPHABET S} Q_{s s_t} (1-p(s_t,u)) 2 \mathcal M(y),
  \label{eq:third-term_M}
\end{align}
where $\mathcal M$ is the cumulative density function of $\mu$ and in equality $(a)$, we use the evenness of $\mu$.
The following lemma shows that $M(y|e,s,u)$ is increasing in $e$ for all $e \in \reals_{\ge 0}$. 
\begin{lemma} \label{lemma:Mu-inc}
  For any $y \in \reals_{\ge 0}$, $M(y|e,s,u)$ is increasing in $e$, $e \in
  \reals_{\ge 0}$. 
\end{lemma}
\begin{proof}
  Let $M_e(y|e,s;u)$ denote $\partial M/\partial e$. Then
  \[
    M_e(y|e,s;u) = \sum_{s_t \in \ALPHABET S} Q_{s s_t} p(s,u) a
    \big[ \mu(y - ae) - \mu(y + ae) \big].
  \]
  From (M0) and Lemma~\ref{lem:mono}, we get that $M_e(y|e,s;u) \ge 0$ for any
  $e,y \in \reals_{\ge 0}$ and $u \in \ALPHABET U$. Thus, $M(y|e,s;u)$ is
  increasing in $e$. 
\end{proof}

\begin{remark}\label{rem:discrete-case}
  For $\ALPHABET X = \integers$, one needs to replace the partial derivatives
  with finite differences. See~\cite{JC-AM-mono-TAC2019} for the proof.
\end{remark}

\subsubsection{Condition~(C4)}
\emph{For any $s \in \ALPHABET S$, $u_1, u_2 \in \ALPHABET U$, $u_1 < u_2$, $c(e,s,u)$ is submodular in $(e,u)$ on $\reals_{\ge 0} \times \ALPHABET U$}. 

Consider $e_1,e_2 \in \reals_{\ge 0}$ such that $e_1 \le e_2$. Then,
\begin{align}
  \hskip 2em & \hskip -2em
  c(e_1,s,u_2) - c(e_2,s,u_2) \notag\\
  &= \sum_{s_t \in \ALPHABET S} Q_{s s_t}p(s_t,u_2) (d(e_1)-d(e_2))\notag\\
  & \stackrel{(a)}{\ge} \sum_{s_t \in \ALPHABET S} Q_{s s_t} p(s_t,u_1) (d(e_1)-d(e_2)) \notag\\
  & = c(e_1,s,u_1) - c(e_2,s,u_1)
  \label{eq:c_t_submod},
\end{align}
where $(a)$ holds since $d(\cdot)$ is $\SQC(0)$, with $d(e) \ge 0$, for all $e \in \reals$ and $p(s_t,\cdot)$ is decreasing. Eq.~\eqref{eq:c_t_submod} implies that for any $s\in \ALPHABET S$, $c(e,s,u)$ is submodular in $(e,u)$ on $\reals_{\ge 0} \times \ALPHABET U$.

\subsubsection{Condition~(C5)}
\emph{For any $s \in \ALPHABET S$, $u \in \ALPHABET U$ and $y \in \reals_{\ge 0}$, $M(y\,|\, e,s;u)$ is submodular in $(e,u)$ on $\reals_{\ge 0} \times \ALPHABET U$}.

It is shown in Lemma~~\ref{lemma:Mu-inc} that $M$ is increasing in $e$. Recall $M_e$ as introduced in the proof of Lemma~\ref{lemma:Mu-inc}. From (M2), $M_e(y|e,s;u)$ is decreasing in $u$. Thus, $M(y|e,s;u)$ is submodular in $(e,u)$ on $\reals_{\ge 0} \times \ALPHABET U$. 

\section{Proof of Property~\ref{prop:6}}
Consider uncountable $\ALPHABET U$ and $\ALPHABET X = \reals$. Similar derivation holds for finite $\ALPHABET U$ and $\ALPHABET X = \integers$. 
\begin{enumerate}
  \item Recall from~\eqref{eq:B01}, 
    \begin{align*}
      B(\pi, s, \varphi) 
      &= \int_\reals \pi(e) p(s, \varphi(e)) de \\
      &=\int_{\ALPHABET U} \Big[\int_\reals \pi(e) \IND_{\{\varphi(e) = u\}} de \Big] p(s,u) du \\
      & \stackrel{(a)}{=} \int_{\ALPHABET U} \Big[\int_\reals \xi(e) \IND_{\{\theta(e) = u\}} de \Big] p(s,u) du\\
      &= B(\xi, s, \theta),
    \end{align*}
    where $(a)$ holds by Remark~\ref{rem:probU}.
  \item Recall from Theorem~\ref{thm:structure-fin},
    \begin{align*}
      \Lambda (\pi, \varphi) &= \int_\reals \lambda(\varphi(e)) \pi(e) de\\
      &= \int_{\ALPHABET U} \Big[\int_\reals \pi(e)\IND_{\{\varphi(e) = u\}} de \Big] \lambda(u) du\\
      & \stackrel{(b)}{=} \int_{\ALPHABET U} \Big[\int_\reals \xi(e) \IND_{\{\theta(e) = u\}} de \Big]\lambda(u) du\\
      & = \Lambda (\xi, \theta),
    \end{align*}
    where $(b)$ holds by Remark~\ref{rem:probU}.
\end{enumerate}

\end{document}